\documentclass[a4paper,12pt,fleqn]{article}
\usepackage[DIV13]{typearea}
\usepackage{amsmath}
\usepackage{amssymb}
\usepackage{amscd}
\usepackage{amsfonts}
\usepackage{bbold}
\usepackage{cite}
\usepackage[footnotesize]{caption2}
\usepackage{graphicx}
\usepackage[center,footnotesize,hang]{subfigure}
\usepackage{url}

\newcommand{\eV}{\ensuremath{\,\mathrm{eV}}}
\newcommand{\keV}{\ensuremath{\,\mathrm{keV}}}
\newcommand{\MeV}{\ensuremath{\,\mathrm{MeV}}}
\newcommand{\GeV}{\ensuremath{\,\mathrm{GeV}}}
\newcommand{\TeV}{\ensuremath{\,\mathrm{TeV}}}

\def\I{\; \mathrm{i}}

\DeclareMathOperator{\diag}{diag}

\newcommand{\Delatm}{\Delta m_{31} ^{2}}
\newcommand{\Delsol}{\Delta m_{21} ^{2}}

\newcommand{\Cl}[1]{\mathcal{C} _{#1}}

\newcommand{\Ord}[2]{\; ^{\circ} \mathrm{#1}_{#2}  \;}
\newcommand{\OrdCl}[1]{\; ^{\circ} \mathcal{C} _{#1} \;}
\newcommand{\Rep}[1]{\underline{\mbox{\textbf{#1}}}}
\newcommand{\MoreRep}[2]{\underline{\mbox{\textbf{#1}}} _{\mbox{\textbf{#2}}}}
\newcommand{\Groupname}[2]{$ {#1} _{#2} $}
\newcommand{\Doub}[2]{$ {#1} _{#2} ^{\prime} $}

\newcommand{\Eqref}[1]{Eq.\eqref{#1}}

\newcommand{\Tabref}[1]{Table \ref{#1}}

\newcommand{\Appref}[1]{Appendix \ref{#1}}

\allowdisplaybreaks[1]

\begin{document}

\begin{titlepage}

\ \vspace*{-15mm}
\begin{flushright}
TUM-HEP-620/06
\end{flushright}
\vspace*{5mm}

\begin{center}
{\Large\sffamily\bfseries
\mathversion{bold} \Groupname{S}{4} \mathversion{normal} Flavor
Symmetry and Fermion Masses: \\ Towards a Grand Unified
Theory of Flavor}
\\[13mm]
{\large
C. Hagedorn\footnote{E-mail: \texttt{chagedor@ph.tum.de}}$^{(a)}$,
M. Lindner\footnote{E-mail: \texttt{lindner@ph.tum.de}}$^{(a)}$ and
R.N. Mohapatra\footnote{E-mail: \texttt{rmohapat@physics.umd.edu}}$^{(b)}$}
\\[5mm]
{\small\textit{$^{(a)}$
Physik-Department T30d, Technische Universit\"{a}t M\"{u}nchen\\
James-Franck-Stra{\ss}e, 85748 Garching, Germany
}}
\\[3mm]
{\small\textit{$^{(b)}$
Department of Physics and Center for String and Particle Theory,\\
University of Maryland, College Park, MD-20742, USA}}
\end{center}
\vspace*{1.0cm}

\begin{abstract}

\noindent Pursuing a bottom-up approach to explore which flavor
symmetry could serve as an explanation of the observed fermion
masses and mixings, we discuss an extension of the standard model (SM)
where the flavor structure for both quarks and leptons is
determined by a spontaneously broken \Groupname{S}{4} and
the requirement that its particle content is embeddable simultaneously into the
conventional $SO(10)$ grand unified theory (GUT) and a continuous flavor
symmetry $G_{f}$ like $SO(3)_{f}$ or $SU(3)_{f}$. We explicitly provide
the Yukawa and the Higgs sector of the model and show its viability in
two numerical examples which arise as small deviations from rank one
matrices. In the first case, the corresponding mass matrix is
democratic and in the second one only
its 2-3 block is non-vanishing. We demonstrate that the Higgs
potential allows for the appropriate vacuum expectation
value (VEV) configurations in both cases, if CP is conserved. For the
first case, the chosen Yukawa couplings can
be made natural by invoking an auxiliary \Groupname{Z}{2} symmetry. The
numerical study we
perform shows that the best-fit values for the lepton mixing angles
$\theta_{12}$ and
$\theta_{23}$ can be accommodated for normal neutrino mass hierarchy. The
results for the quark
mixing angles turn out to be too small. Furthermore the CP-violating
phase $\delta$ can only be
reproduced correctly in one of the examples. The small mixing angle values
are likely to be brought into the experimentally allowed ranges by
including radiative corrections. Interestingly, due to the
\Groupname{S}{4} symmetry the mass matrix of the right-handed
neutrinos is proportional to the unit matrix.

\end{abstract}

\end{titlepage}

\setcounter{footnote}{0}

\section{Introduction}
\label{sec:Intro}

The discovery of neutrino masses and attempts to understand the
flavor puzzle have made it quite natural to expect the existence
of an embedding product group for the SM such as $SO(10) \times G
_{f}$ \cite{so10gfcont,so10gfdisc} at very high energies, where
$SO(10)$ acts as gauge and $G _{f}$ as a flavor (family) symmetry.
All fermions of one generation can then be unified into the spinor
representation $\Rep{16}$ of $SO(10)$ and the three known
generations are assigned to representations of $G_{f}$. To specify
$G_{f}$ two important properties have to be fixed: $G_{f}$ can be
either abelian or non-abelian and it can be either continuous or
discrete. Abelian symmetries are not able to explain the existence
of more than one generation, since their representations are all
one-dimensional. Therefore we consider such a choice of $G_{f}$ to
be less interesting than the others, although many models have
obtained interesting results from $U(1)$ flavor  symmetries
\cite{u1gut}. Most of the other models in which $G_{f}$ is
non-abelian \cite{so10gfcont} have some common features, like
additional $U(1)$ or \Groupname{Z}{n} factors, heavy vector-like
fermions, elaborate mechanisms for the VEV alignment, to make them
viable. Further unification might explain these assumptions.
 We search for a simpler model by first
constructing a low energy theory with the SM gauge group and a
discrete non-abelian $G_{f}$ and then showing possible embeddings
of this theory into an $SO(10)$ GUT and a continuous $G_{f}$. To
really unify all three generations $G_{f}$ should be either $SO(3)
_{f}$ or $SU(3) _{f}$. Our discrete symmetry will therefore be a
subgroup of $SO(3)$ or $SU(3)$. We like to point out that none of
the models mentioned in reference \cite{so10gfcont} with $G_{f}$
being continuous follows this strategy. In cases where $G_{f}$ is
discrete \cite{so10gfdisc}, the symmetry is always broken at the
GUT scale so that these models are also not comparable to our
ansatz. Models in which the discrete non-abelian flavor symmetry
is only broken at low energies became very popular in the last few
years \cite{s3,a4,s4,s4so10,dns,dnprimes,tprime,deltas}. However,
these models can rarely be embedded into $SO(10) \times G_{f}$.

\noindent In working with discrete symmetries two issues are necessary
to mention. First, the breaking of a discrete global symmetry does not
lead to (unwanted) massless Goldstone bosons,
unlike continuous symmetries. Second, if this breaking is only
spontaneous, it might produce domain walls which can be a serious problem
\cite{domain}. It can be solved by either invoking low scale
inflation or embedding the discrete symmetry into a continuous gauge
symmetry \cite{domainsol}. A further issue which is only important, if
the symmetry should be gauged, is the question of anomalies. Since we
intend to embed the discrete symmetry into a continuous one at high
energies, it is enough to make sure that the continuous one is
anomaly-free. This can be checked by calculating the usual triangle
diagrams. If the symmetry turns out to be anomalous, adding
appropriate representations can solve this problem.

\noindent Searching for an adequate discrete group we concentrate
on the smallest subgroups of $SO(3)$ or $SU(3)$ which have at
least one irreducible three-dimensional representation.
\Groupname{A}{4} is the smallest of these groups and has already
been discussed extensively in the literature \cite{a4}. The second
smallest group, sometimes called \Groupname{T}{7}, is rarely
known, but its properties are in some sense similar to
\Groupname{A}{4}. Concerning the three-dimensional representations
the same holds for the groups \Doub{T}{} \cite{tprime} (\Doub{T}{}
two-valued group of \Groupname{T}{} and \Groupname{T}{} isomorphic
to \Groupname{A}{4}) and \Groupname{T}{h} (\Groupname{T}{h}
isomorphic to \Groupname{A}{4} $\times$ \Groupname{C}{2}). In this
paper, we focus on the group \Groupname{S}{4}, which has a
different group structure and is therefore worth exploring as a
possible flavor symmetry. This has already been done in several
papers \cite{s4,s4so10}.  Pakvasa et al. \cite{s4} constructed an SM-like model for
quarks whose transformation properties under \Groupname{S}{4}, however, do neither allow
an embedding into $SO(10)$ nor into $SO(3) _f$ ($SU(3) _f$). An
$SO(10)$ model presented by one of the authors (R.N.M.) and Lee
\cite{s4so10} uses similar \Groupname{S}{4} representations for
the fermions and Higgs fields, but leads to the SMA solution to
the solar neutrino problem which is ruled out \footnote{The
renormalization group extrapolations of neutrino masses in this
model have not been studied and they might make a difference to
the above conclusion since neutrinos are quasi-degenerate in this
case.}. Recently, also Ma \cite{s4} discussed an \Groupname{S}{4}
model, but he only considered the neutrino sector. Therefore
there is still no working \Groupname{S}{4} model which addresses
the flavor structure of all fermions in a unified manner and can
be embedded into a GUT as well as a continuous flavor symmetry.
The aim of our study is to fill this gap.

\indent As noted, in this paper, we consider the non-abelian
discrete flavor symmetry \Groupname{S}{4} accompanying the SM
gauge group which naturally commutes with it. The gauge and the
flavor symmetry are assumed to be both broken only spontaneously
at the electroweak scale.  The particle content consists of the
three known fermion generations together with three right-handed
neutrinos and a set of six Higgs fields being SM doublets and
transforming as $\MoreRep{1}{1} + \Rep{2} + \MoreRep{3}{1}$ under
\Groupname{S}{4}. The choice of the fermion content allows the
embedding of our model into $SO(10) \times G _{f}$. We observe
that the number of parameters determining the mass matrices in our
model, i.e. the Yukawa couplings and the VEVs of the Higgs fields,
equals the number of observables, i.e. masses and mixing angles,
in the CP-conserving case. If CP is violated, there are more
parameters in the model than observables to fit. However, it is
rather non-trivial to find parameter configurations which can
reproduce all data, since \Groupname{S}{4} constrains the mass
structures as well as the Higgs potential. In our numerical study
this turns out to be possible except for rather small deviations of the
quark mixings which might be fixed by radiative corrections.

\indent The paper is organized as follows: Section 2 contains
the group theory of \Groupname{S}{4}, in section 3 we
present our model and argue why this is the minimal model with
respect to the possible embeddings into $SO(10)$ and $G _{f}$. Then,
we display the structure of the mass matrices
arising from the \Groupname{S}{4} invariant Yukawa couplings.
After setting our conventions for the mixing matrices we give some
numerical examples and comment on their viability. To complete the
model, we calculate the Higgs potential and perform a restricted
analysis of its possible minima. In section 6 we conclude and
point to open questions in our model. Finally, the appendices
contain Kronecker products, Clebsch Gordan coefficients and embeddings
of \Groupname{S}{4} as well as some information on the Higgs potential
and the selected minima.

\mathversion{bold}
\section{The \Groupname{S}{4} Group}
\mathversion{normal}
\label{sec:S4group}

The group \Groupname{S}{4} is the permutation group of four
distinct objects. It is isomorphic to the group
\Groupname{O}{} which is the symmetry group of a regular octahedron and so well-known in solid
state physics. Its order is 24, i.e. it has 24 distinct
elements. \Groupname{S}{4} has five conjugate classes and therefore contains five irreducible
representations which are all real. Among these are two
one-dimensional ones, the identity
(i.e. the representation being invariant under all transformations of
\Groupname{S}{4}, also called the symmetric representation) and the anti-symmetric one (i.e. the one changing
sign under odd permutations, also called alternating). In the following we will
denote the identity one with $\MoreRep{1}{1}$ and the anti-symmetric
one with $\MoreRep{1}{2}$. There is one two-dimensional representation
called $\Rep{2}$ and two three-dimensional ones, $\MoreRep{3}{1}$ and
$\MoreRep{3}{2}$. Out of these five irreducible representations only the two
three-dimensional ones are faithful. Their characters $\chi$,
i.e. the traces of
their representation matrices, are given in the
character table, see \Tabref{chartabS4}. There we use the following notations: $\Cl{i}$ with
$i=1,...,5$ are the five classes of the
group, $\OrdCl{i}$ is the order of the $i ^{\mathrm{th}}$ class,
i.e. the number of distinct elements contained in this class, $\Ord{h}{\Cl{i}}$
is the order of the elements $R$ in the class $\Cl{i}$, i.e. the smallest
integer ($>0$) for which the equation $R ^{\Ord{h}{\Cl{i}}}= \mathbb{1}$
holds. Furthermore the table contains one representative for each
class $\Cl{i}$ given as product of the generators $\rm
A$ and $\rm B$ of the group.
\begin{table}
\begin{center}
\begin{tabular}{|l|ccccc|}
\hline
&\multicolumn{5}{|c|}{classes}                                                 \\ \cline{2-6}
&$\Cl{1}$&$\Cl{2}$&$\Cl{3}$&$\Cl{4}$&$\Cl{5}$\\
\cline{1-6}
\rule[0.15in]{0cm}{0cm} $\rm G$         &$\rm \mathbb{1}$&$\rm A^{2}$
&$\rm A B^{2}$ &$\rm B$ &$\rm A $\\
\cline{1-6}
$\OrdCl{i}$                  &1      &3      &6      &8      &6\\
\cline{1-6}
$\Ord{h}{} _{\Cl{i}}$                           &1      &2      &2      &3      &4\\
\hline
\rule[0.4cm]{0.3cm}{0cm}$\MoreRep{1}{1}$                                &1      &1      &1      &1      &1              \\[0.1cm]
\rule[0cm]{0.3cm}{0cm}$\MoreRep{1}{2}$                                &1      &1      &-1     &1      &-1             \\[0.1cm]
\rule[0cm]{0.3cm}{0cm}$\Rep{2}$                                       &2      &2      &0      &-1     &0              \\[0.1cm]
\rule[0cm]{0.3cm}{0cm}$\MoreRep{3}{1}$                                &3      &-1     &1      &0      &-1             \\[0.1cm]
\rule[0cm]{0.3cm}{0cm}$\MoreRep{3}{2}$                                &3      &-1     &-1     &0      &1              \\[0.1cm]
\hline
\end{tabular}
\end{center}
\begin{center}
\begin{minipage}[t]{12cm}
\caption[]{ Character table of the group
  \Groupname{S}{4}. (explanation see text) \label{chartabS4}}
\end{minipage}
\end{center}
\end{table}
\noindent From the generators $\rm A$ and $\rm B$ all other elements
of \Groupname{S}{4} can be formed by multiplication. They ought to fulfill the following
relations \cite{Lomont} :
\begin{equation}
\label{generatoreq}
\rm A^4= \mathbb{1} \;\; , \;\;  B^3=\mathbb{1} \;\;\; \mbox{and} \;\;\;
 AB^2 A=B \;\; , \;\; ABA=BA^2B \; .
\end{equation}
\noindent We show one possible choice of generators in \Appref{app:repmats}. Using them we calculate the Clebsch Gordan
coefficients for all the Kronecker products.

\noindent \Groupname{S}{4} is the smallest group containing one-, two-
and three-dimensional representations together with the group
\Doub{T}{}.

\noindent \Groupname{S}{4} can be embedded into $SO(3)$ as well as
in $SU(3)$ (where it is isomorphic to the group $\Delta (24)$
\cite{su3subgroups}) and therefore gives the opportunity to embed
our discrete flavor symmetry into a continuous one which is broken
at a high energy scale. Possible embedding schemes are shown in
\Appref{app:embeds}.

\noindent In our model the group \Groupname{S}{4} is broken
completely at the electroweak scale, however this breaking could
also occur in two steps such that \Groupname{S}{4} breaks to one
of its subgroups which is then completely broken. The non-abelian
subgroups of  \Groupname{S}{4} turn out to be already well-known
as flavor symmetries: they are \Groupname{S}{3} (which is
isomorphic to \Groupname{D}{3}), \Groupname{D}{4} and
\Groupname{A}{4}. Correlation tables containing the corresponding
breaking sequences for the representations of \Groupname{S}{4} can
be found in \cite{Califano}.

\mathversion{bold}
\section{The \Groupname{S}{4} Model}
\mathversion{normal}
\label{sec:model}

\subsection{Particle Assignment}

In this subsection, we describe how to assign the fermions and Higgs
bosons to different \Groupname{S}{4} representations in such a way that
the model can be embedded simultaneously into $SO(10)$ and a
continuous flavor symmetry $G _{f}$. We argue that our choice
which is displayed in \Tabref{particletab} is unique in that sense.

\noindent To embed the model into $SO(10)$ all the fermion generations
have to transform in the same way under \Groupname{S}{4}.
Furthermore they have to transform either as trivial or as fundamental
representation of the flavor group $G _{f}$, since all the
other representations of these groups
have a dimension larger than three. The only choice is then that
they transform as $\MoreRep{3}{2}$ under \Groupname{S}{4} apart
from the trivial one where all the generations just form total
singlets under the flavor group. An important point to note is that
it is not possible to assign the three generations to
$\MoreRep{3}{1}$, since this representation cannot be identified
with the fundamental representation of $SO(3)_{f}$ or $SU(3)_{f}$.
Therefore our assignment is unique.

\vspace{0.05in}

\noindent In the next step, we have to choose the representations
for the Higgs fields in order to give masses to the SM fermions. This
choice depends on the desired Yukawa structure as well as
the constraint to fill (a) certain representation(s) of the
embedding group $G _{f}$. For fermions transforming as
$\MoreRep{3}{2}$ under \Groupname{S}{4}, the Higgs fields which
can couple in an \Groupname{S}{4} invariant manner belong either to
$\MoreRep{1}{1}$, $\Rep{2}$, $\MoreRep{3}{1}$ or
to $\MoreRep{3}{2}$, i.e. the only representation which cannot
couple to form a total singlet under \Groupname{S}{4} is
$\MoreRep{1}{2}$. Taking now for simplicity only the couplings
which are symmetric in flavor space, i.e. lead to symmetric mass
matrices, one is left with the representations $\MoreRep{1}{1}$,
$\Rep{2}$ and $\MoreRep{3}{1}$. Regarding the possible embeddings
into $SO(3)_{f}$ one recognizes that at least five Higgs fields
transforming as $\Rep{2} + \MoreRep{3}{1}$ are needed and for an
embedding into $SU(3)_{f}$ one needs six fields $\sim
\MoreRep{1}{1} + \Rep{2} + \MoreRep{3}{1}$. Furthermore it turns
out that the minimal version of five Higgs fields $\sim \Rep{2} +
\MoreRep{3}{1}$ is not phenomenologically viable, since it leads to
traceless symmetric mass
matrices. Therefore the minimal setup of Higgs fields that we
choose to get fermion masses contains six fields transforming as
$\MoreRep{1}{1} + \Rep{2} + \MoreRep{3}{1}$ under
\Groupname{S}{4}. In the case of an embedding into $SO(3) _{f}$ these
representations are identified with $\Rep{1} + \Rep{5}$ and in the
case of $SU(3) _{f}$ they are unified into the six-dimensional
representation of $SU(3) _{f}$ (see: \Appref{app:embeds}). 

\noindent Neutrinos can also have Majorana masses apart from Dirac
mass terms. Since the three right-handed neutrinos are also unified into one
$\MoreRep{3}{2}$ under \Groupname{S}{4} the only invariant mass
term for the right-handed neutrinos is simply proportional to the
unit matrix. The embedding of our model into $SO(3) _{f}$ does not change
this, however one has to keep in mind that in $SU(3)_{f}$ the irreducible
three-dimensional representation with which $\MoreRep{3}{2}$ of
\Groupname{S}{4} is identified is complex and therefore does not allow
an invariant direct mass term for the right-handed neutrinos. Keeping
our model as minimal as possible its embedding into $SO(3) _{f}$
instead of $SU(3) _{f}$ is therefore preferred. Nevertheless, the inclusion of gauge singlets transforming
as $\Rep{6}$ under $SU(3)_{f}$ can give masses to the right-handed
neutrinos, if the singlets acquire an appropriate VEV. Since the six-dimensional representation of $SU(3)_{f}$
contains a total singlet of \Groupname{S}{4} (see \Appref{app:embeds}),
this does not necessarily lead to the breaking of \Groupname{S}{4} at
a high energy scale. Furthermore notice that the situation changes, if the
model is embedded into a GUT like $SO(10)$ at the same time. This will
be discussed below.

\noindent A non-trivial structure for the right-handed Majorana mass term
requires the introduction of gauge singlets which transform
non-trivially under \Groupname{S}{4}. To implement the canonical
type I seesaw \cite{seesaw1} the VEVs of such fields ought to be
of the order of $10^{13} \GeV$. In this case, the flavor symmetry
\Groupname{S}{4} is broken at this high energy scale rather than the
electroweak scale. Hence we discard this possibility. To keep the
model as minimal as possible we also do not include $SU(2) _{L}$
Higgs triplets which could give rise to a type II seesaw mass for
the light neutrinos \cite{seesaw2}. In some classes of
$SO(10)$ models this has become quite popular in order to get a relation
between the light neutrino mass matrix and the difference of the
down and charged lepton mass matrix
\cite{so10massreltype2seesaw}. Nevertheless fits using type I seesaw
in $SO(10)$ can also be found \cite{so10type1seesaw}.

\noindent To embed the model as a whole into $SO(10)$, the Higgs
fields also have to be identified with certain $SO(10)$
representations. In order to get a tree-level coupling they have
to be part of either $\Rep{10}$, $\Rep{120}$ or
$\overline{\Rep{126}}$ under $SO(10)$. Since we fixed the Yukawa
couplings to be symmetric in flavor space the representation
$\Rep{120}$ drops out. Furthermore we observe that we need one
$\overline{\Rep{126}}$ which transforms trivially under
\Groupname{S}{4} for the right-handed Majorana mass term. We ought
to choose its VEVs such that it does not give rise to a mass term
for the left-handed neutrinos, since $SU(2) _{L}$ Higgs triplets
are absent in our low energy model. The minimal choice of fields
would be: six Higgs fields transforming as $(\Rep{10},
\MoreRep{1}{1} + \Rep{2} + \MoreRep{3}{1})$\\ under $(SO(10),
\mbox{\Groupname{S}{4}})$ and one $\overline{\Rep{126}} \sim
\MoreRep{1}{1}$ for the mass term of the right-handed neutrinos.
The $\overline{\Rep{126}}$ should also contribute to the Dirac
mass term of the other fermions, since otherwise the masses of the
down quarks and charged leptons were the same. In our opinion this
is still not enough. Therefore at best one promotes each SM Higgs
doublet to one $\Rep{10}$ and one $\overline{\Rep{126}}$, such
that the $SO(10)$ model has six ten- and six 126-dimensional Higgs
representations. Among the $\overline{\Rep{126}}$s only the one
which transforms trivially under \Groupname{S}{4} should develop a
VEV at high energies and the rest only at the electroweak scale.

\noindent To complete the model one needs at least
one further Higgs representation, for example a $\Rep{210}$. This
representation together with the \Groupname{S}{4} invariant 126-dimensional representation
should break $SO(10)$ down to the SM with the Pati-Salam
group as intermediate group. However it should not break the flavor group
\Groupname{S}{4} and hence has to be assigned to $\MoreRep{1}{1}$
under \Groupname{S}{4}. With all these Higgs fields we believe that it is
possible to make a viable high energy completion of our low energy
model.

\noindent In the numerical examples given below we fit the masses and
mixing angles at the scale $\mu$ which equals the $W$ boson mass. In
order to perform a fit at the GUT scale instead we would have to take into
account the renormalization group running of all masses and couplings
which will be complicated in a model with such a rich Higgs structure.

\noindent Some issues still need to be discussed. Without constructing
the $SO(10)$ invariant Higgs potential it remains the question whether
the advocated VEV configuration can be realized. Since our model
contains several ten- and 126-dimensional representations and at
least one 210-dimensional one, there is a doublet-doublet splitting
problem, i.e. one has to ensure that only six of the SM-like Higgs
doublets have masses at the electroweak scale while the others acquire
masses around the GUT scale. In the same manner one has to solve the
well-known doublet-triplet splitting problem. In general separating
the electroweak and the GUT scale will be difficult without
supersymmetry (SUSY). Furthermore it has to be guaranteed that the gauge
couplings unify at all being at the same time still in the
perturbative regime at the GUT scale.

\noindent  Finally, the $SO(10) \times \mbox{\Groupname{S}{4}}$ model is embedded into
$SO(10) \times G_{f}$. As already mentioned, the three generations of
fermions are identified with the fundamental representation of
$G_{f}$. The Higgs fields contained in the six ten-dimensional
representations of $SO(10)$ are unified into the two representations
$(\Rep{10},\Rep{1})$ and $(\Rep{10},\Rep{5})$ for $G_{f}= SO(3) _{f}$
and into $(\Rep{10},\Rep{6})$ for $G_{f}=SU(3) _{f}$. The Higgs fields
in the $\overline{\Rep{126}}$s are treated in a similar way. Therefore the
right-handed neutrinos acquire masses by coupling to
$(\overline{\Rep{126}},\Rep{1})$ in case of $G_{f}$ being $SO(3)_{f}$ while they
couple to $(\overline{\Rep{126}},\Rep{6})$, if
$G_{f}$ is $SU(3)_{f}$. The $\Rep{210}$ needed for the gauge symmetry
breaking transforms trivially under $G_{f}$ in the simplest case. Furthermore we need
some $SO(10)$ gauge singlets breaking
$G_{f}$ in such a way that only \Groupname{S}{4} remains. The
smallest non-trivial possibility for $G_{f} = SO(3) _{f}$ is
$\Rep{9}$ and for $G_{f} = SU(3) _{f}$ $\Rep{6}$, since they
contain an \Groupname{S}{4} singlet (see \Appref{app:embeds}). But
this we will not consider any further in this paper.

\noindent To sum up, our model now contains three generations of
fermions all transforming as $\MoreRep{3}{2}$ and six SM-like Higgs
fields transforming as $\MoreRep{1}{1} + \Rep{2} + \MoreRep{3}{1}$
under \Groupname{S}{4}. In the next subsection, we show the arising
Dirac and Majorana mass matrices.
\begin{table}
\begin{center}
\begin{tabular}{|c|c|c|}\hline
Particle & $SU(3)_c \times SU(2)_L\times U(1)_Y$ rep. &
\Groupname{S}{4} rep.\\
\hline
Quarks $Q$ & $(\Rep{3},\Rep{2}, +{1\over 3})$ & $\MoreRep{3}{2}$\\
 Antiquarks $u^c$ &  $(\overline{\Rep{3}},\Rep{1}, - {4\over 3})$ & $\MoreRep{3}{2}$\\
Antiquarks  $ d^c$ &  $(\overline{\Rep{3}}, \Rep{1},+\frac{2}{3})$ & $\MoreRep{3}{2}$\\
 Leptons $L$ & $(\Rep{1}, \Rep{2} -1)$ & $\MoreRep{3}{2}$\\
 Antileptons  $e^c$ & $(\Rep{1},\Rep{1},+2)$ & $\MoreRep{3}{2}$\\
 Right-handed $\nu$s  $\nu^c$ & $(\Rep{1},\Rep{1},0)$ & $\MoreRep{3}{2}$\\
\hline
Doublet Higgs  $\bf \phi_0$ & $(\Rep{1}, \Rep{2}, -1)$ & $\MoreRep{1}{1}$\\
Doublet Higgs $\bf (\phi_1,\phi_2)$ & $(\Rep{1},\Rep{2}, -1)$ & $\Rep{2}$\\
Doublet Higgs $\bf (\xi_1,\xi_2,\xi_{3})$ & $(\Rep{1}, \Rep{2}, -1)$ & $\MoreRep{3}{1}$\\
\hline
\end{tabular}
\end{center}
\begin{center}
\begin{minipage}[t]{12cm}
\caption[]{ The particle content and its symmetry properties
under \Groupname{S}{4}.\label{particletab}}
\end{minipage}
\end{center}
\end{table}
\subsection{Fermion Masses}
\label{sec:chargedmasses} The \Groupname{S}{4} invariant Yukawa
couplings in our model are
\small
\begin{eqnarray} \nonumber
\mathcal{L}_{Y} &=& \alpha^{u} _{0} \, (Q_{1} \, u^c + Q_{2} \, c^{c} +
Q_{3} \, t^{c}) \, \tilde{\phi}_{0} + \alpha^{u}_{1} \, (\sqrt{3} \, (Q_{2} \,
c^c - Q_{3} \, t^c) \, \tilde{\phi}_{1} + (-2 \, Q_{1} \, u^c + Q_{2} \, c^c +
Q_{3} \, t^c) \, \tilde{\phi}_{2}) \\ \nonumber
&+& \alpha^{u} _{2} \, ((Q_{2} \, t^c + Q_{3} \, c^c) \tilde{\xi}_{1} + (Q_{1}
\, t^c + Q_{3} \, u^c) \, \tilde{\xi}_{2} + (Q_{1} \, c^c + Q_{2}  \, u^c) \,
\tilde{\xi}_{3}) \\ \nonumber
 &+& \alpha^{d} _{0} \, (Q_{1} \, d^c + Q_{2} \, s^{c} +
Q_{3} \, b^{c}) \, \phi_{0} + \alpha^{d}_{1} \, (\sqrt{3} \, (Q_{2} \,
s^c - Q_{3} \, b^c) \, \phi_{1} + (-2 \, Q_{1} \, d^c + Q_{2} \, s^c +
Q_{3} \, b^c) \, \phi_{2}) \\ \nonumber
&+& \alpha^{d} _{2} \, ((Q_{2} \, b^c + Q_{3} \, s^c) \xi_{1} + (Q_{1}
\, b^c + Q_{3} \, d^c) \, \xi_{2} + (Q_{1} \, s^c + Q_{2}  \, d^c) \,
\xi_{3}) \\ \nonumber
&+& \alpha^{e} _{0} \, (L_{1} \, e^c + L_{2} \, \mu^{c} +
L_{3} \, \tau^{c}) \, \phi_{0} + \alpha^{e}_{1} \, (\sqrt{3} \, (L_{2} \,
\mu^c - L_{3} \, \tau^c) \, \phi_{1} + (-2 \, L_{1} \, e^c + L_{2} \, \mu^c +
L_{3} \, \tau^c) \, \phi_{2}) \\ \nonumber
&+& \alpha^{e} _{2} \, ((L_{2} \, \tau^c + L_{3} \, \mu^c) \xi_{1} + (L_{1}
\, \tau^c + L_{3} \, e^c) \, \xi_{2} + (L_{1} \, \mu^c + L_{2}  \, e^c) \,
\xi_{3}) \\ \nonumber
&+& \alpha^{\nu} _{0} \, (L_{1} \, \nu_{e}^c + L_{2} \, \nu_{\mu}^{c} +
L_{3} \, \nu_{\tau}^{c}) \, \tilde{\phi}_{0} + \alpha^{\nu}_{1} \, (\sqrt{3} \, (L_{2} \,
\nu_{\mu}^c - L_{3} \, \nu_{\tau}^c) \, \tilde{\phi}_{1} + (-2 \, L_{1} \, \nu_{e}^c + L_{2} \, \nu_{\mu}^c +
L_{3} \, \nu_{\tau}^c) \, \tilde{\phi}_{2}) \\
&+& \alpha^{\nu} _{2} \, ((L_{2} \, \nu_{\tau}^c + L_{3} \, \nu_{\mu}^c) \tilde{\xi}_{1} + (L_{1}
\, \nu_{\tau}^c + L_{3} \, \nu_{e}^c) \, \tilde{\xi}_{2} + (L_{1} \,
\nu_{\mu}^c + L_{2}  \, \nu_{e}^c) \, \tilde{\xi}_{3})
\end{eqnarray}
\normalsize
where the fields $\tilde{\phi}_{0,1,2}$ and $\tilde{\xi}_{1,2,3}$ are
the conjugates of the fields $\phi_{0,1,2}$ and $\xi_{1,2,3}$ related
by  $\tilde{\phi} = \epsilon \, \phi ^{\star}$ with $\epsilon$ being
the 2-by-2 anti-symmetric matrix in $SU(2)_{L}$ space and the star
denotes the complex conjugation.

\noindent They lead to the following mass matrices for $i= u,d,e, \nu$:
\begin{equation}
\label{massmatrix}
\mathcal{M}^{i}=
\left(
  \begin{array}{ccc}
    \alpha^{i}_{0} \, \phi_{0} - 2 \, \alpha^{i}_{1}
    \, \phi_{2} & \alpha^{i}_{2} \, \xi_{3} &
    \alpha^{i}_{2} \, \xi_{2} \\
    \alpha^{i}_{2} \, \xi_{3} & \alpha^{i}_{0} \,
    \phi_{0} + \alpha^{i}_{1} \, (\sqrt{3} \, \phi_{1} +
    \phi_{2}) & \alpha^{i}_{2} \, \xi_{1} \\
    \alpha^{i}_{2} \, \xi_{2} & \alpha^{i}_{2} \, \xi_{1}
    & \alpha^{i}_{0} \, \phi_{0}+ \alpha^{i}_{1} \, (-\sqrt{3} \, \phi_{1} +
    \phi_{2})
\end{array}
\right)
\end{equation}
with the Higgs fields being replaced by their VEVs for the down quarks
and the charged leptons and by the complex conjugate of their VEVs for
the up quarks and the neutrinos. The sum of the VEVs has to be equal the
electroweak scale, i.e. $\sum \limits _{i} ^{}
|\mbox{\scriptsize{VEV}}_{i}|^2 \approx (174 \GeV)^2$. Note that the
fields $\phi_{0,1,2}$ only appear in the diagonal entries. The
contribution coming from the Higgs field $\phi_{0}$ is proportional to
the unit matrix, since $\phi_{0}$ transforms trivially under
\Groupname{S}{4}. The fields $\phi_{1}$ and $\phi_{2}$ on the other
hand are coupled in such a manner that their contribution is
traceless. Finally, the fields $\xi_{i}$ which form a triplet under
\Groupname{S}{4} only induce flavor-changing interactions, i.e. their
contributions are encoded in the off-diagonal elements of the mass
matrix $\mathcal{M} ^{i}$. Generally all the parameters in
\Eqref{massmatrix} can be complex.

\vspace{0.05in}

\noindent In the case of CP-conservation, we arrive at twelve real
Yukawa couplings, five real VEVs for the Higgs doublet fields and the
right-handed neutrino mass scale $M_{R}$ (see below). The
sixth VEV is fixed by the electroweak scale. The 18 couplings and VEVs
correspond to the twelve masses for the quarks, the charged leptons and the light neutrinos, the three
CKM mixing angles and the three leptonic mixing angles
whereof two have been measured. With CP-violation - either explicit
or spontaneous - the number of parameters is increased such that it
exceeds the number of observables. Nevertheless, it is not apparent
that the mass structure restricted by the \Groupname{S}{4} symmetry
allows one to fit all the data. Therefore, we perform a numerical
study in the next section.

\noindent It is interesting to note that assigning the fermion
generations to $\MoreRep{3}{1}$ instead of $\MoreRep{3}{2}$ leads
to exactly the same mass matrices, but does not allow an
embedding into $G _{f}$  without adding at least
two further chiral generations to complete a representation of $G _{f}$ (see \Appref{app:embeds}).

\vspace{0.05in}

\noindent Next we discuss the neutrino sector. Since the right-handed
neutrinos transform as $\MoreRep{3}{2}$ under
\Groupname{S}{4}, their mass matrix $M_{RR}$ is
proportional to the unit matrix, i.e. $M_{RR} = M_{R}
\mathbb{1}$. This means that the mass matrix for the
light neutrinos arising from the type I seesaw has the form:
\begin{equation}
\label{lightnumatrix}
M_{\nu} = (-) \frac{1}{M_{R}} \mathcal{M}^{\nu}
\mathcal{M}^{\nu \, T} = (-) \frac{1}{M_{R}} (\mathcal{M}^{\nu})^2 \; .
\end{equation}
The last step is allowed, since all the Dirac mass matrices are
symmetric by construction.

\noindent The fact $M_{RR} \propto \mathbb{1}$ indicates that the seesaw
mechanism cannot be the sole origin of the difference between the
quark and the lepton mixings in our model. As we will see below, in
some cases there
exists the possibility to impose an
additional symmetry to maintain the diverse mixing patterns.

\noindent With regard to leptogenesis, $M_{RR} \propto \mathbb{1}$ is a viable
starting point for the mechanism of resonant
leptogenesis \cite{reslepto}, since small radiative corrections can
generate the small mass splittings which are needed.

\noindent Finally , we want to indicate how the mass matrices change
in case of a full $SO(10)$ model. First the number of Yukawa couplings
is reduced from twelve $\left\{ \alpha ^{i} _{j} \right\}$ with
$i=u,d,e, \nu$ and $j= 0,1,2$ to only six $\left\{ \alpha _{j \, R}
\right\}$ ($j=0,1,2$ and $R=\Rep{10}$ or $R=\overline{\Rep{126}}$), i.e. three couplings to
$(\Rep{10}, \MoreRep{1}{1})$, $(\Rep{10}, \Rep{2})$ and $(\Rep{10},
\MoreRep{3}{1})$ and three to $(\overline{\Rep{126}}, \MoreRep{1}{1})$,
$(\overline{\Rep{126}}, \Rep{2})$ and $(\overline{\Rep{126}},\MoreRep{3}{1})$. At the
same time the number of VEVs is in general increased. 

\noindent The additional embedding of \Groupname{S}{4} into a continuous
flavor symmetry $G _{f}$ reduces the number of Yukawa couplings
to its minimum. In case of $G_{f}$ being $SO(3) _{f}$ the three
Yukawa couplings $\left\{ \alpha _{j \, R} \right\}$ for each $R =
\Rep{10}$ or $R = \overline{\Rep{126}}$ can be expressed as two
independent ones: $\alpha _{0 \, R} \equiv \alpha _{R}$ and
$\alpha _{1 \, R} = \alpha _{2 \, R} \equiv \beta _{R}$, since the
two- and three-dimensional representation under \Groupname{S}{4}
will be unified into one five-dimensional one. For $G _{f} = SU(3)
_{f}$ we are left with only one Yukawa coupling for each $R =
\Rep{10}$ or $R = \overline{\Rep{126}}$, since the six-dimensional
representation of $SU(3) _{f}$ breaks up into $\MoreRep{1}{1} +
\Rep{2} + \MoreRep{3}{1}$ under \Groupname{S}{4}. 

\noindent At the end, taking this setup does not reduce the number of parameters, but they
can be eventually correlated such that predictions can be made.

\section{Phenomenology of the Mass Structures}
\label{sec:numerics}

In this section we show that our model allows viable solutions. For this 
purpose, we introduce our conventions for the mixing matrices
and give the experimental data.

\subsection{Conventions for the Mixing Matrices}
\label{sec:convmix}

The Dirac mass matrices arise from the coupling:
\[
y_{ij} \, \mathrm{L}^{T} _{i} \, \epsilon \, \phi \, \mathrm{L}^{\mathrm{c}} _{j}
\]
for the downtype quarks ($\mathrm{L}=Q$,
$\mathrm{L}^{\mathrm{c}}=d^{c}$) and charged leptons ($\mathrm{L}=L$,
$\mathrm{L}^{\mathrm{c}}=e^{c}$) and for the uptype ones ($\mathrm{L}=Q$,
$\mathrm{L}^{\mathrm{c}}=u^{c}$)
and the neutrinos ($\mathrm{L}=L$, $\mathrm{L}^{\mathrm{c}}=\nu^{c}$) :
\[
y_{ij} \, \mathrm{L}^{T} _{i} \, \epsilon \, \tilde{\phi} \, \mathrm{L}^{\mathrm{c}} _{j}
\]
\noindent The mass matrices for the quarks are diagonalized by:
\[
U_{u} ^{\dagger} \mathcal{M} ^{u} \mathcal{M} ^{ u \, \dagger} U_{u}=
\diag(m_u ^2, m_c ^2, m_t ^2) \; , \;\; U_{d} ^{\dagger} \mathcal{M} ^{d}
\mathcal{M} ^{d \, \dagger} U_{d}= \diag(m_d ^2, m_s ^2, m_b ^2)
\]
where $U_{u}$ and $U_{d}$ are the unitary matrices transforming the
left-handed up and down quarks to their mass eigenstates.
The CKM mixing matrix is given by:
\[
V_{CKM}= U_{u} ^{T} \, U_{d} ^{\star} \; .
\]
The standard parameterization for $V_{CKM}$ is \cite{PDG}:
\[
V_{CKM}= \left( \begin{array}{ccc}
c_{12} c_{13} & s_{12} c_{13} & s_{13} \, e^{-i \delta}
\\[0.2cm]
-s_{12} c_{23} - c_{12} s_{23} s_{13} \, e^{i \delta}
& c_{12} c_{23} - s_{12} s_{23} s_{13}\, e^{i \delta}
& s_{23} c_{13} \\[0.2cm]
s_{12} s_{23} - c_{12} c_{23} s_{13} \, e^{i \delta}
&
- c_{12} s_{23} - s_{12} c_{23} s_{13}\, e^{i \delta}
& c_{23} c_{13}\\
\end{array}  \right)
\]
where we use the abbreviations $s_{ij}= \sin (\theta_{ij})$ and
$c_{ij} = \cos (\theta_{ij})$. The angles are restricted to lie in the
first quadrant and $\delta$ can take any value between 0 and $2 \, \pi$.\\
Similarly, in the leptonic sector the mass matrix for the charged
leptons fulfills the relation:
\[
U_{l} ^{\dagger} \mathcal{M} ^{l} \mathcal{M} ^{l \, \dagger} U_{l} = \diag(m_e ^2, m_{\mu} ^2, m_{\tau} ^2)
\]
with the unitary matrix $U_{l}$ transforming the left-handed charged
leptons to their mass eigenstates.
\noindent The light neutrino mass matrix $M_{\nu}$ coming from the type I
seesaw is a complex symmetric matrix. Hence it can be diagonalized by
$U_{\nu}$:
\[
U_{\nu} ^{\dagger} M_{\nu} U_{\nu}^{\star} = \diag
(m_1,m_2,m_3) \; , \;\; U_{\nu} ^{\dagger} M _{\nu} M_{\nu} ^{\dagger} U_{\nu} = \diag(m_1 ^2, m_2 ^2, m_3 ^2)
\]
with the mass eigenvalues $m_{i}$ being positive definite.
The $U_{MNS}$ matrix is defined as
\[
\nu_{\alpha \, L}= \sum \limits _{i=1} ^{3} U_{MNS} ^{\alpha \, i} \, \nu _{i
  \, L} \;\;\; \mbox{for} \;\;\; \alpha= e, \, \mu, \, \tau \;\;\;
\mbox{and} \;\;\; i=1,2,3
\]
where the $\nu_{\alpha \, L}$ denote the flavor and the $\nu _{i \,
  L}$ the mass eigenstates.\\
Therefore $U_{MNS}$ is expressed as
\[
U_{MNS}= U_{l} ^{T} \, U_{\nu} ^{\star} \; .
\]
\noindent In the case of Majorana neutrinos the $U_{MNS}$ matrix can
be factorized in a unitary matrix which is  parameterized  in the same
way as the CKM matrix $V_{CKM}$ and a diagonal matrix containing the
two Majorana phases $\varphi_{1}$ and $\varphi_{2}$.
\[
U_{MNS} = \tilde{V}_{CKM} \cdot \diag(\rm{e} ^{\I \varphi_1}, \rm{e}
^{\I \varphi_2}, 1) \; .
\]
Both Majorana phases are taken to fulfill $0 \leq \varphi_{1}, \varphi_{2}
\leq \pi$.

\subsection{Experimental Data}

The quark masses at the scale $\mu \approx M_{W}$ are given by  \cite{masses}:
\[
m_u = 2.2 \MeV \; , \;\; m_c =0.81 \GeV \; , \;\; m_t = 170 \GeV \; ,
\]
\[
m_d = 4.4 \MeV \; , \;\; m_s = 80 \MeV \; , \;\; m_b = 3.1 \GeV \; .
\]
The mixing angles and the CP-phase measured in tree-level processes
only are \cite{PDG}:
\[
s_{12} =0.2243 \pm 0.0016 \; , \;\; s_{23}= 0.0413 \pm 0.0015 \; , \;\;
s_{13}= 0.0037 \pm 0.0005 \; , \;\; \delta= 1.05 \pm 0.24 \; .
\]
They are almost independent of the scale $\mu$ at low energies. To
quantify the CP-violation one can introduce the Jarlskog invariant
$\mathcal{J}_{CP}$ \cite{jarlskog}:
\[
\mathcal{J} _{CP}= \left( 2.88 \pm 0.33 \right) \, \times \, 10^{-5}
\; .
\]
\noindent The charged lepton masses at $\mu \approx M_{W}$ are:
\[
m_e= 511 \keV \; , \;\; m_{\mu}= 106 \MeV \; , \;\; m_{\tau}= 1.78
\GeV \; .
\]
In the neutrino sector only the two mass squared differences measured
in atmospheric and solar neutrino experiments are known \cite{nufit}:
\[
\Delsol = m_2 ^2 -m_1 ^2 = (7.9^{+ 0.6} _{- 0.6} ) \times 10^{-5} \eV
^2 \; , \;\; |\Delatm|
= |m_3 ^2 -m_1 ^2| = (2.2 ^{+ 0.7} _{-0.5} ) \times 10^{-3}  \eV^2 \; .
\]
The leptonic mixing angles are constrained by experiments:
\[
s_{13} ^{2} \leq 0.031  \; , \;\; s_{12} ^{2}= 0.3 ^{+ 0.04} _{-0.05}
\; , \;\; s_{23} ^{2}= 0.5 ^{+ 0.14} _{-0.12} \; .
\]
All values observed in neutrino oscillations are given at $2 \,
\sigma$ level. The three possible CP-phases $\delta$, $\varphi_{1}$
and $\varphi_{2}$ in the leptonic sector have not been measured till today.

\subsection{Numerical Examples}

In this section, we present two numerical examples which can
reconcile the data apart from the fact that some of the
quark mixings turn out to be smaller than the central
values.

\noindent They correspond to
perturbations around two different rank one matrices for the
quarks and charged leptons:
\begin{equation}
\mathcal{M} _{1} = \left( \begin{array}{ccc}
                             1 & 1 & 1\\
                             1 & 1 & 1\\
                             1 & 1 & 1
                             \end{array}
                   \right)          \;\;\; \mbox{and} \;\;\;
\mathcal{M} _{2} =  \left( \begin{array}{ccc}
                             0 & 0 & 0\\
                             0 & 1 & 1\\
                             0 & 1 & 1
                             \end{array}
                   \right) \; ,
\end{equation}
i.e. the matrix $\mathcal{M} _{1}$ is the democratic mass matrix,
which has been discussed several times in the literature \cite{democraticmass}, and
$\mathcal{M} _{2}$ only has a non-vanishing 2-3
block. Matrices of the form of $\mathcal{M} _{2}$ are often used
as approximation for the light neutrino mass matrix to produce
maximal atmospheric mixing \cite{23blockmass}. Both matrices are able to explain the
strong hierarchies observed in the quark and the charged lepton
sector. For the neutrinos it is necessary that their
mass matrix significantly deviates from the forms of $\mathcal{M}
_{1}$ and $\mathcal{M} _{2}$ to satisfy the restrictions on the
lepton mixing matrix. In particular in our two numerical examples
the neutrinos dominantly couple to the Higgs fields $\phi_{1,2}$,
i.e.
 their mass matrix has large entries on its diagonal.

\noindent In the following we show that certain VEV configurations
which reflect our flavor symmetry together with some fine-tuning of
the Yukawa couplings allow the matrices $\mathcal{M}_{1}$ and
$\mathcal{M}_{2}$ to result from the general form given in \Eqref{massmatrix}.

\noindent First, we consider the matrix $\mathcal{M} _{1}$. In the CP-conserving case, the
Higgs potential has a minimum at which the VEVs
of the fields $\xi_{i}$ are equal, the VEVs of $\phi_{1,2}$ are
zero and $\phi_{0}$ has a non-vanishing VEV. The VEVs of $\xi_{i}$
and the field $\phi_{0}$ are not necessarily equal, but we make
the natural assumption that they are almost the same. With this
and the constraint that the Yukawa couplings $\alpha _{0} ^{i}$
and $\alpha _{2} ^{i}$ should be the same, the Dirac mass matrices
$\mathcal{M} ^{i}$ have the democratic form. We then perturb
around this known minimum of the Higgs potential and also allow
changes in the Yukawa couplings $\alpha_{j} ^{i}$ in order to get
mass matrices whose masses and mixing parameters agree with the
observed ones. For concreteness:
\begin{eqnarray}\nonumber
\alpha_{0} ^{u} &=& 0.651341 - 0.00001 \I \; , \;\; \alpha_{1} ^{u} =
-0.0058575 + 0.001286 \I \; , \;\; \alpha_{2} ^{u}= 0.651341 \; , \\ \nonumber
\alpha_{0} ^{d} &=& 0.011598 + 0.000219 \I  \; , \;\; \alpha_{1} ^{d} =
-0.00299585 -0.00098905 \I \; , \\ \nonumber \alpha_{2} ^{d} &=& 0.011708 -
0.000284 \I \; , \\ \nonumber
\alpha_{0} ^{e} &=&  0.0071585 - 2.0 \cdot 10^{-7} \I \; , \;\;
\alpha_{1} ^{e} = -0.000375685 + 0.00331405 \I \; , \;\; \alpha_{2}
^{e}= 0.0065433 \; , \\ \nonumber
\alpha_{0} ^{\nu} &=& 0.15224 + 0.0906 \I \; , \;\; \alpha_{1} ^{\nu}
= 1.04426 -0.018245 \I \; , \;\; \alpha_{2} ^{\nu}= 0.08364 + 0.04734
\I \; , \\ \nonumber
\langle \phi_{0} \rangle &=& \left( 87 + 0.01776 \I \right) \, \GeV \; , \;\; \langle
\phi_{1} \rangle = \left( -2.5912 - 14.1982 \I \right) \, \GeV \; , \\
\nonumber \langle \phi_{2} \rangle &=& \left( 1.27 - 7.6236 \I \right)
\, \GeV \; , \;\; \langle \xi_{1} \rangle = \left( 86.8607 + 1.14585
  \I \right) \, \GeV \; , \\ \nonumber 
\langle \xi_{2} \rangle &=& \left( 87 - 0.11162 \I \right) \, \GeV \; , \;\; \langle \xi_{3} \rangle =
\left( 87 + 0.88384 \I \right) \, \GeV \; , \\ \nonumber
%|\langle \phi_{0} \rangle |^2 &+& |\langle \phi_{1} \rangle |^2 + |\langle \phi_{2} \rangle |^2 + |\langle \xi_{1} \rangle |^2 + |\langle \xi_{2} \rangle | ^2 + |\langle \xi_{3} \rangle | ^2 \approx (174.705 \GeV)^2 \\ \nonumber
M_{R} &=& 4.3 \, \times \, 10^{13} \GeV \; .
\end{eqnarray}
The hierarchies among the Yukawa couplings $\alpha ^{i} _{0} \sim \alpha _{2} ^{i} \gg \alpha _{1} ^{i}$ for $i=u,d,e$ and $\alpha ^{\nu} _{1} \gg \alpha _{2} ^{\nu} \sim \alpha _{0} ^{\nu}$ can nicely be explained by an approximate auxiliary \Groupname{Z}{2} under which only the right-handed neutrinos and the Higgs fields $\phi_{1}$ and $\phi_{2}$ transform:
\begin{equation}
\nu ^{c} _{i} \;\;\; \rightarrow \;\;\; - \nu ^{c} _{i} \;\;\; \mbox{and} \;\;\; \phi_{1,2} \;\;\; \rightarrow \;\;\; - \phi_{1,2}
\end{equation}
whereas the rest remains invariant.
The structure of the resulting mass matrices is then democratic for
the quarks and the charged leptons while it is dominated by the
$(2,2)$ entry for the light neutrinos. The numerical values given here
lead to:
\begin{eqnarray}\nonumber
m_{u} &=& 2.2 \MeV \; , \;\; m_{c} = 0.814 \GeV \; , \;\; m_{t}=
169.94 \GeV \; , \\ \nonumber
m_{d} &=& 4.46 \MeV \; , \;\; m_{s} = 81.2 \MeV \; , \;\; m_{b} = 3.05
\GeV \; ,\\ \nonumber
m_{e} &=& 514 \keV \; , \;\; m_{\mu} = 105.5 \MeV \; , \;\; m_{\tau} =
1.76 \GeV \; , \\ \nonumber
\Delatm &=& 2.3 \, \times \, 10^{-3} \eV ^2 \; , \;\; \Delsol = 7.89 \,
\times \, 10^{-5} \eV^2 \; .
\end{eqnarray}
The sum of the three light neutrino masses is $\sum _{i} \; m_{i} =
0.0614 \eV$, i.e. the mass spectrum is strongly hierarchical with the
smallest mass $m_{1} \approx 0.0038 \eV$.
% m1= 0.0038 , m2= 0.0097, m3=0.0479
This is well below the mass
bounds known from cosmology: $\sum _{i} m_{i} < (0.42  \dots 1.8)
\eV$. This upper bound depends on whether the measurements of the Lyman $\alpha$ spectrum
are included or not \cite{cosmomassbounds}.

\noindent The quark mixing angles turn out to be
\[
s_{12}= 0.2238 \; , \;\; s_{13}= 0.003694 \; , \;\; s_{23} = 0.02831
\; .
\]
Unfortunately, $s_{23}$ is too small. This may be
compensated by radiative corrections. The CP-phase $\delta$ is $1.174$ radian
and therefore near the upper bound $1.29$ radian. The Jarlskog
invariant has the value $\mathcal{J} _{CP} = 2.1 \, \times \, 10^{-5}$. More
interesting to see is that the neutrino mixing angles can be
accommodated:
\[
s_{12} ^{2}= 0.3 \; , \;\; s_{23} ^2 = 0.49946 \; ,
\]
and we predict $|U ^{e3} _{MNS}|$ to be $0.06616$. This is within
the reach of the next generation experiments
\cite{theta13prospect}. Furthermore, the three leptonic CP-phases are:
\[
\delta = 0.9983 \; , \;\; \varphi_{1} = 1.579 \; , \;\;
\varphi_{2}= 1.336 \;\; \mbox{in radian} .
\]
\noindent Calculating for completeness the magnitudes of $|m_{ee}|$
and $m_{\beta}$ which can be extracted from neutrinoless double beta
decay and beta decay, respectively, one finds:
\[
|m_{ee}|= |\sum \limits _{i=1} ^{3} \left( U_{MNS} ^{e \, i} \right)
^{2} \,  m_{i}| =
0.0054 \eV \;\;\; \mbox{and} \;\;\; m_{\beta} = \left( \sum \limits _{i=1}
^{3} |U_{MNS} ^{e \, i}| ^{2} \, m_{i} ^{2} \right) ^{1/2} = 0.0069 \eV
\]
\noindent These values are at least two orders of magnitude below the current
experimental bounds which are $|m_{ee}| \leq 0.9 \eV$ \cite{0vbbbounds} and $m_{\beta}
\leq 2.2 \eV$ \cite{betadecaybounds}. They are also below the limits of the experiments
planned for the next years \cite{0vbbprospect,betadecayprospect}. The
main reason for this is the strong hierarchy in the neutrino mass spectrum.

\noindent If we take the second mass matrix $\mathcal{M} _{2}$ as
starting point, we find the following numerical example:
\begin{eqnarray}\nonumber
\alpha_{0} ^{u} &=& 0.56672 - 0.00001 \I \; , \;\; \alpha_{1} ^{u} =
0.2833 + 0.00001 \I \; , \;\; \alpha_{2} ^{u}= 0.85175 + 0.00608 \I \;
,\\ \nonumber
\alpha_{0} ^{d} &=& 0.01028 - 3 \cdot 10^{-6} \I  \; , \;\; \alpha_{1}
^{d} = 0.005145 - 0.000021 \I \; , \;\; \alpha_{2} ^{d} = 0.015597 +
0.000789 \I \; ,\\ \nonumber
\alpha_{0} ^{e} &=& 0.0059333 - 9 \cdot 10^{-6} \I \; , \;\;
\alpha_{1} ^{e} = 0.0029676 - 10^{-6} \I \; , \;\; \alpha_{2} ^{e}=
0.0088571 + 0.0010664 \I \; ,\\ \nonumber
\alpha_{0} ^{\nu} &=& 0.00333 + 0.028 \I \; , \;\; \alpha_{1} ^{\nu} =
0.51567 -0.01616 \I \; , \;\; \alpha_{2} ^{\nu}= -0.01572 + 0.89947 \I
\; ,\\ \nonumber
\langle \phi_{0} \rangle &=& \left( 99.9974 + 0.0026 \I \right) \,
\GeV \; , \;\; \langle
\phi_{1} \rangle = \left( -2.67385 - 6.79332 \I \right) \, \GeV \; ,
\\ \nonumber 
\langle \phi_{2} \rangle &=& \left( 99.9923 - 0.02867 \I  \right) \,
\GeV \; , \;\; \langle \xi_{1} \rangle = \left( 99.9907 -
  0.266 \I \right) \, \GeV \; , \\ \nonumber \langle \xi_{2} \rangle &=&
\left( 0.0058 - 0.38774 \I \right) \, \GeV \; , \;\; \langle \xi_{3}
\rangle = \left( 0.04332 - 0.14671 \I \right) \, \GeV \; ,\\ \nonumber
%|\langle \phi_{0} \rangle |^2 &+& |\langle \phi_{1} \rangle |^2 + |\langle \phi_{2} \rangle |^2 + |\langle \xi_{1} \rangle |^2 + |\langle \xi_{2} \rangle | ^2 + |\langle \xi_{3} \rangle | ^2 \approx (173.348 \GeV)^2 \\ \nonumber
M_{R} &=& 6.3 \, \times \, 10^{13} \GeV \; .
\end{eqnarray}
Here, one can clearly see that the Yukawa couplings have to
fulfill  the relation: $\alpha_{0} ^{i}: \alpha _{1} ^{i} : \alpha
_{2} ^{i} \approx 2:1:3$ for $i=u,d,e$ to produce a matrix with a
dominant 2-3 block. The resulting masses and mass squared
differences are given by:
\begin{eqnarray}\nonumber
m_{u} &=& 2.4 \MeV \; , \;\; m_{c} = 0.812 \GeV \; , \;\; m_{t}=
170.24 \GeV \; ,\\ \nonumber
m_{d} &=& 4.4 \MeV \; , \;\; m_{s} = 80 \MeV \; , \;\; m_{b} = 3.10
\GeV \; ,\\ \nonumber
m_{e} &=& 512.6 \keV \; , \;\; m_{\mu} = 106 \MeV \; , \;\; m_{\tau} =
1.78 \GeV \; ,\\ \nonumber
\Delatm &=& 2.4 \, \times \, 10^{-3} \eV^2 \; , \;\; \Delsol = 7.59 \,
\times \, 10^{-5} \eV^2 \; .
\end{eqnarray}
The mass spectrum for the three light neutrinos is normally
ordered and degenerate with $m_{1} \approx 0.1682 \eV$ and $\sum
_{i} m_{i} = 0.512 \eV$ which coincides with the upper bound of
the cosmological measurements, if the Lyman $\alpha$ data is also
taken into account.
% m1= 0.1682 , m2= 0.1684, 0.1752
For the quarks all three mixing angles are
about $10 \%$ too small:
\[
s_{12}= 0.2128 \; , \;\; s_{13}= 0.0038 \; , \;\; s_{23} = 0.0389 \; .
\]
Again, this has to be compensated by radiative corrections. The
CP-phase $\delta$ turns out to be the more severe problem, since
$\delta \approx 0.386$ radian, i.e. the CP-violation generated
by this setup is more than a factor of two too small and so is the
Jarlskog invariant $\mathcal{J} _{CP} = 1.15 \, \times \, 10^{-5}$. The two
measured mixing angles in the lepton sector can be adjusted to
their currently given best-fit values, i.e. $s_{12} ^2 = 0.306$
and $s_{23} ^2 = 0.506$. The third mixing
angle $\theta_{13}$ and the three leptonic CP-phases turn out to be:
\[
s_{13}^2 = 0.0034  \;\; (|U_{MNS}^{e3}|=0.0584)\; , \;\; \delta = 3.032
\; , \;\; \varphi_{1} = 3.102 \; , \;\; \varphi_{2} = 3.081 \; ,
\]
Again, all phases are given in radian. $|m_{ee}|$ and $m_{\beta}$ have
almost the same value $0.168 \eV$, since the neutrino masses are nearly
degenerate and all phases are approximately $\pi$. $m_{\beta}$ is near the
limit which will be reached by the KATRIN experiment
\cite{betadecayprospect}. Due to the degeneracy of the neutrino mass
spectrum also $|m_{ee}|$ \cite{0vbbprospect} and the
sum of the neutrino masses \cite{cosmomassprospect} can be measured by the
up-coming experiments in the next five to ten years.

\vspace{0.05in}

\noindent The matrix structure is very similar to the one of $\mathcal{M}
_{2}$ for the quarks as well as for the charged
leptons while the matrix for the light neutrinos has approximately
the form:
\begin{equation}
|M _{\nu}| \sim \left( \begin{array}{ccc}
            2 & 0 & 0\\
            0 & 1 & 2\\
            0 & 2 & 1
    \end{array}
    \right)  \; .
\end{equation}
In this case one also finds that the invoked VEV configuration
$\langle \phi_{1} \rangle = \langle \xi_{2} \rangle = \langle \xi_{3}
\rangle = 0$ together with the other VEVs being nearly the same is an
allowed minimum of the Higgs potential, if CP is conserved.

\noindent In the two examples shown here we have taken all
parameters to be complex, i.e. we have assumed explicit CP-violation
and have not made use of possible field re-definitions to absorb some
of the complex phases. We have done so in order to keep as many (free)
parameters as possible in the numerical fit procedure. For example,
this does not mean that there is no viable
solution in the case of spontaneous CP-violation.

\section{The Higgs Potential and its Possible Minima}

Finally, we discuss the \Groupname{S}{4} invariant Higgs potential $V$
and show
that there exist two CP-conserving minima from which the VEV
configurations assumed above can arise as ``perturbations''.
\small
\begin{eqnarray}
\label{Higgspot}
V~&=&~-\mu^2_1 \, (\phi^\dagger_0\phi_0)-\mu^2_2 \, \sum \limits
_{j=1} ^{2} \phi^\dagger_j \phi_j - \mu^2_{3} \, \sum \limits _{i=1} ^{3}
\xi^{\dagger}_{i} \xi_{i} \\ \nonumber
 &+& \lambda_0 \, (\phi^\dagger_0\phi_0)^2 +\lambda_1 \, (\phi^\dagger_1\phi_1+
\phi^\dagger_2\phi_2)^2~ + \lambda_2 \, (\phi^\dagger_1 \phi_2-
\phi^\dagger_2 \phi_1)^2\\ \nonumber
&+&\lambda_3 \left[(\phi^\dagger_1\phi_2+
\phi^\dagger_2\phi_1)^2 + (\phi^\dagger_1\phi_1-
\phi^\dagger_2\phi_2)^2\right] \\ \nonumber
&+& \sigma_{1} \, (\phi_0^\dagger \phi_0)(\phi^\dagger_1\phi_1+
\phi^\dagger_2\phi_2)+\left\{\sigma_{2} \left[ (\phi^\dagger_0\phi_1)^2+
(\phi^\dagger_0\phi_2)^2 \right] + \mathrm{h.c.} \right\}\\  \nonumber
&+& \tilde{\sigma}_{2} \, \left[|\phi^\dagger_0 \phi_1|^2+|\phi^\dagger_0 \phi_2|^2 \right]
+\left\{ \sigma_{3} \, \left[ (\phi^\dagger_0 \phi_1 )(\phi^\dagger_1 \phi_2+
  \phi^\dagger_2 \phi_1) + (\phi^\dagger_0 \phi_2) (\phi^\dagger_1
  \phi_1-\phi^\dagger_2 \phi_2) \right]
+ \mathrm{h.c.} \right\} \\ \nonumber
&+& \lambda^{\xi}_{1} \, \left( \sum \limits _{i=1} ^{3}
\xi^{\dagger}_{i} \xi_{i} \right) ^2 + \lambda^{\xi} _{2} \, \left[ 3 \,
(\xi^{\dagger}_{2} \xi_{2} - \xi^{\dagger} _{3} \xi_{3})^2 + (-2 \,
\xi^{\dagger}_{1} \xi_{1} + \xi_{2} ^{\dagger} \xi_{2} + \xi_{3}
^{\dagger} \xi_{3})^2 \right] \\ \nonumber
&+& \lambda^{\xi} _{3} \, \left[ (\xi_{2} ^{\dagger} \xi_{3} + \xi_{3}
^{\dagger} \xi_{2})^2 + (\xi_{1} ^{\dagger} \xi_{3} + \xi_{3}
^{\dagger} \xi_{1})^2 +(\xi_{1} ^{\dagger} \xi_{2} + \xi_{2}
^{\dagger} \xi_{1})^2 \right] \\ \nonumber
&+& \lambda^{\xi} _{4} \, \left[ (\xi_{2} ^{\dagger} \xi_{3} - \xi_{3}
^{\dagger} \xi_{2})^2 + (\xi_{1} ^{\dagger} \xi_{3} - \xi_{3}
^{\dagger} \xi_{1})^2 +(\xi_{1} ^{\dagger} \xi_{2} - \xi_{2}
^{\dagger} \xi_{1})^2 \right] \\ \nonumber
&+& \tau_{1} \, (\phi^{\dagger} _{0} \phi_{0}) \left( \sum \limits _{i=1}
^{3} \xi^{\dagger}_{i} \xi_{i} \right) +  \tau_{2} \, \left( \sum \limits _{j=1}
^{2} \phi^{\dagger} _{j} \phi_{j} \right) \left( \sum \limits _{i=1}
^{3} \xi^{\dagger}_{i} \xi_{i} \right) \\ \nonumber
&+& \tau_{3} \, \left[ \sqrt{3} \, (\phi_{1} ^{\dagger} \phi_{2} + \phi_{2}
^{\dagger} \phi_{1})(\xi_{2} ^{\dagger} \xi_{2} - \xi_{3} ^{\dagger}
\xi_{3}) + (\phi_{1} ^{\dagger} \phi_{1} - \phi_{2} ^{\dagger}
\phi_{2})(-2 \, \xi^{\dagger} _{1} \xi_{1} + \xi^{\dagger} _{2}
\xi_{2} + \xi_{3} ^{\dagger} \xi_{3}) \right] \\ \nonumber
&+& \left\{ \kappa_{1} \, \left[ 4 \, (\phi_{2} ^{\dagger} \xi_{1})^2 +
  (\sqrt{3} \, \phi_{1} ^{\dagger} \xi_{2} + \phi_{2} ^{\dagger}
  \xi_{2})^2 + (\sqrt{3} \, \phi_{1} ^{\dagger} \xi_{3} - \phi_{2}
  ^{\dagger} \xi_{3})^2 \right] + \mathrm{h.c.} \right\} \\ \nonumber
&+& \tilde{\kappa}_{1}
\, \left[ 4 \, |\phi_{2} ^{\dagger} \xi_{1}|^2 +
  |\sqrt{3} \, \phi_{1} ^{\dagger} \xi_{2} + \phi_{2} ^{\dagger}
  \xi_{2}|^2 + |\sqrt{3} \, \phi_{1} ^{\dagger} \xi_{3} - \phi_{2}
  ^{\dagger} \xi_{3}|^2 \right] \\ \nonumber
&+& \left\{ \kappa_{2} \, \left[ 4 \, (\phi_{1} ^{\dagger} \xi_{1})^2 +
  (\sqrt{3} \, \phi_{2} ^{\dagger} \xi_{2} - \phi_{1} ^{\dagger}
  \xi_{2})^2 + (\sqrt{3} \, \phi_{2} ^{\dagger} \xi_{3} + \phi_{1}
  ^{\dagger} \xi_{3})^2 \right] + \mathrm{h.c.} \right\} \\ \nonumber
&+& \tilde{\kappa}_{2}
\, \left[ 4 \, |\phi_{1} ^{\dagger} \xi_{1}|^2 +
  |\sqrt{3} \, \phi_{2} ^{\dagger} \xi_{2} - \phi_{1} ^{\dagger}
  \xi_{2}|^2 + |\sqrt{3} \, \phi_{2} ^{\dagger} \xi_{3} + \phi_{1}
  ^{\dagger} \xi_{3}|^2 \right] \\ \nonumber
&+& \left\{ \kappa_{3} \, \left[ 2 \, (\phi_{2} ^{\dagger} \xi_{1})(\xi_{2}
  ^{\dagger} \xi_{3} + \xi_{3} ^{\dagger} \xi_{2}) - (\sqrt{3} \,
  \phi_{1} ^{\dagger} \xi_{2} + \phi_{2}^{\dagger} \xi_{2})(\xi_{1} ^{\dagger}
  \xi_{3} + \xi_{3} ^{\dagger} \xi_{1}) + (\sqrt{3} \, \phi_{1}
  ^{\dagger} \xi_{3} - \phi_{2} ^{\dagger} \xi_{3})(\xi_{1} ^{\dagger}
  \xi_{2} + \xi_{2} ^{\dagger} \xi_{1}) \right] \right.
\\ \nonumber &+& \left.   \mathrm{h.c.} \right\}  \\ \nonumber
&+&  \left\{ \kappa_{4} \, \left[ 2 \, (\phi_{1} ^{\dagger} \xi_{1})(\xi_{3}
  ^{\dagger} \xi_{2} - \xi_{2} ^{\dagger} \xi_{3}) + (\sqrt{3} \,
  \phi_{2} ^{\dagger} \xi_{2} - \phi_{1}^{\dagger} \xi_{2})(\xi_{1} ^{\dagger}
  \xi_{3} - \xi_{3} ^{\dagger} \xi_{1}) - (\sqrt{3} \, \phi_{2}
  ^{\dagger} \xi_{3} + \phi_{1} ^{\dagger} \xi_{3})(\xi_{2} ^{\dagger}
  \xi_{1} - \xi_{1} ^{\dagger} \xi_{2}) \right] \right. \\ \nonumber
&+& \left. \mathrm{h.c.} \right\} \\
\nonumber
&+& \left\{ \kappa_{5} \, \left[ (\phi^{\dagger} _{0} \xi_{1})^2 + (\phi_{0}
  ^{\dagger} \xi_{2})^2 + (\phi_{0} ^{\dagger} \xi_{3})^2 \right] +
  \mathrm{h.c.} \right\} + \tilde{ \kappa}_{5} \, \left[ |\phi^{\dagger} _{0} \xi_{1}|^2 + |\phi_{0}
  ^{\dagger} \xi_{2}|^2 + |\phi_{0} ^{\dagger} \xi_{3}|^2 \right] \\
  \nonumber
&+& \left\{ \kappa_{6} \, \left[ (\phi^{\dagger} _{0} \xi_{1})(\xi_{2}
  ^{\dagger} \xi_{3} + \xi_{3} ^{\dagger} \xi_{2}) + (\phi^{\dagger}
  _{0} \xi_{2})(\xi_{1} ^{\dagger} \xi_{3} + \xi_{3} ^{\dagger}
  \xi_{1}) + (\phi_{0} ^{\dagger} \xi_{3})(\xi_{1} ^{\dagger} \xi_{2}
  + \xi_{2} ^{\dagger} \xi_{1} ) \right] + \mathrm{h.c.} \right\} \\
\nonumber
&+& \left\{ \omega_{1} \, \left[ \sqrt{3} \, (\phi_{0} ^{\dagger}
  \phi_{1})(\xi^{\dagger} _{2} \xi_{2} - \xi^{\dagger} _{3} \xi_{3}) +
  (\phi^{\dagger} _{0} \phi_{2})(-2 \, \xi^{\dagger} _{1} \xi_{1} +
  \xi_{2} ^{\dagger} \xi_{2} + \xi^{\dagger} _{3} \xi_{3}) \right] +
  \mathrm{h.c.} \right\} \\ \nonumber
&+& \left\{ \omega_{2} \, \left[ 2(\phi_{0} ^{\dagger} \xi_{1})(\phi_{2}
  ^{\dagger} \xi_{1}) - (\phi_{0} ^{\dagger} \xi_{2})(\sqrt{3} \,
  \phi_{1} ^{\dagger} \xi_{2} + \phi_{2} ^{\dagger} \xi_{2}) +
  (\phi_{0} ^{\dagger} \xi_{3})(\sqrt{3} \, \phi_{1} ^{\dagger}
  \xi_{3} - \phi_{2} ^{\dagger} \xi_{3}) \right] + \mathrm{h.c.} \right\} \\
\nonumber
&+& \left\{ \omega_{3} \, \left[ 2(\xi_{1} ^{\dagger} \phi_{0})(\phi_{2}
  ^{\dagger} \xi_{1}) - (\xi_{2} ^{\dagger} \phi_{0})(\sqrt{3} \,
  \phi_{1} ^{\dagger} \xi_{2} + \phi_{2} ^{\dagger} \xi_{2}) +
  (\xi_{3} ^{\dagger} \phi_{0})(\sqrt{3} \, \phi_{1} ^{\dagger}
  \xi_{3} - \phi_{2} ^{\dagger} \xi_{3}) \right] + \mathrm{h.c.} \right\}
\end{eqnarray}
\normalsize
where parameters with $+\mathrm{h.c.}$ in curly brackets are in general
complex (for example the last parameter $\omega_{3}$) and the rest is
real. The Higgs potential has 30 parameters in total. 27 of them are
quartic couplings out of which 11 are complex. Interesting to notice,
the potential is invariant under the following transformation:
\begin{equation}
\phi_{1} \;\;\; \rightarrow \;\;\; - \phi_{1} \;\;\; \mbox{and} \;\;\; \xi_{2} \;\;\;
\leftrightarrow \;\;\; \xi_{3}
\end{equation}
and the fields $\phi_{0}$, $\phi_{2}$  and $\xi_{1}$ remain unchanged.

\noindent We can parameterize all possible real VEVs as:
\begin{eqnarray}
\langle \phi_{0} \rangle &=& v_0 \;\; , \;\;\;\;\;\;\;\;\;\;
\langle \phi_{1} \rangle = u \, \cos(\alpha) \;\; , \;\;\;\;\;\;\;\;\;\;
\langle \phi_{2} \rangle = u \, \sin(\alpha) \;\; , \\ \nonumber
\langle \xi_{1} \rangle &=& v \, \cos (\beta) \; , \;\,
\langle \xi_{2} \rangle = v \, \sin (\beta) \, \cos (\gamma) \; , \;\,
\langle \xi_{3} \rangle = v \, \sin (\beta) \, \sin (\gamma) \; .
\end{eqnarray}
The potential at the minimum has then the following form:
\small
\begin{eqnarray}
V_{min} &=& - \mu_{1} ^{2} \, v_0 ^2 - \mu_{2} ^{2} \, u^2 - \mu_{3} ^2
\, v^2 +
\lambda_{0} \, v_0 ^4 + \left(\lambda_{1}+\lambda_{3} \right) \, u^4 \\ \nonumber
&+& \left( \lambda_{1} ^{\xi} + 4 \, \lambda_{3} ^{\xi} \, \sin^2
  (\beta) \right) \, v^4 +
\lambda_{2} ^{\xi} \, \left[ \left(2 - 3 \, \sin^2 (\beta) \right)^2 + 3 \, \sin^4
(\beta) \, \cos^2 (2 \, \gamma) \right] \, v^4  \\ \nonumber &-&
\lambda_{3} ^{\xi} \,\left( 3 + \cos^2 (2 \, \gamma) \right) \, \sin^4 (\beta) \, v^4 + \left( \sigma_{1} + 2
\, \mathrm{Re}(\sigma_{2}) + \tilde{\sigma}_{2} \right) \, v_0 ^2 \, u^2 + 2
\, \mathrm{Re} (\sigma_{3}) \, \sin(3 \alpha) \, v_0 \, u^3 \\
\nonumber &+& \left( \tau_{1} + 2 \, \mathrm{Re}(\kappa_{5}) +
\tilde{\kappa}_{5} \right) \, v_0 ^2 \, v^2 +  \left( 4 \, \mathrm{Re}(\kappa_1 +
\kappa_2) + 2 \, (\tilde{\kappa}_1 + \tilde{\kappa}_2) + \tau_2 \right) \,
u^2 \, v^2  \\ \nonumber &+& \left( 2
\, \mathrm{Re} (\kappa_{1} - \kappa_2)+ \tilde{\kappa}_{1} -
\tilde{\kappa}_2 + \tau_3 \right) \, \left[- \cos (2 \, \alpha) \,
\left( 2 - 3 \, \sin^2 (\beta) \right) \right. \\ \nonumber &+&
\left. \sqrt{3} \, \sin (2 \, \alpha) \, \sin^2
(\beta) \, \cos (2 \, \gamma) \right] \, u^2 \, v^2 \\ \nonumber &+& 3 \,
\mathrm{Re}(\kappa_6) \, \sin(\beta) \, \sin(2 \, \beta) \, \sin (2 \,
\gamma) \, v_0 \, v^3 \\ \nonumber &+& 2 \, \mathrm{Re}
(\omega_{2} + \omega_{3} - \omega_{1}) \left[ \sin
(\alpha) \, \left( 2 - 3 \, \sin^2 (\beta) \right) - \sqrt{3} \, \cos(\alpha) \,
\sin^2 (\beta) \, \cos (2 \gamma) \right] \, u \, v_0 \, v^2
\end{eqnarray}
\normalsize
Note that the couplings $\lambda_{2}$, $\lambda_{4} ^{\xi}$,
$\kappa_{3}$ , $\kappa_{4}$ do not appear in $V_{min}$. One can deduce the following VEV conditions:
\begin{subequations} \label{mincond}
\begin{eqnarray}
\frac{\partial V_{min}}{\partial \alpha} &=& 2 \, u \, \left[ v^2 v_0
  \left( \cos
  (\alpha) \left( 2 -3 \, \sin^2 (\beta) \right) + \sqrt{3} \,  \sin
  (\alpha) \sin ^2 (\beta) \, \cos (2 \, \gamma) \right) \, y_1 \right. \\  \nonumber &+&
\left. u \, v^2 \, \left( \sin (2 \, \alpha) \, \left( 2 - 3 \, \sin^2
      (\beta) \right)  + \sqrt{3} \cos (2 \, \alpha) \sin ^2 (\beta) \cos
  (2 \, \gamma) \right) \, y_2 \right. \\
 \nonumber  &+& \left. 3
 \, v_0 \, u^2 \, \cos (3 \, \alpha) \, \mathrm{Re}(\sigma_{3}) \right]
\end{eqnarray}
\begin{eqnarray}
\frac{\partial V_{min}}{\partial \beta} &=& v^2 \, \left[ -2 \,
  \sqrt{3} \, u \, v_0 \,
\left( \sqrt{3} \, \sin (\alpha) + \cos (\alpha) \, \cos (2 \, \gamma) \right) \,
\sin (2 \, \beta) \, y_1  \right. \\ \nonumber &+& \left. \sqrt{3} \, u^2
\, \left(\sqrt{3} \, \cos (2 \, \alpha) +  \sin (2 \, \alpha) \, \cos(2
  \, \gamma) \right) \, \sin (2 \, \beta) \, y_2  \right. \\
\nonumber &+& \left. 2 \, v^2 \, \left(\sin (4 \, \beta) + \sin^2
    (\beta) \sin (2 \, \beta) \sin^2 (2 \, \gamma) \right) \, y_3
\right. \\
\nonumber &+& \left.
\frac{3}{2} \, v \, v_0 \, \left( 3 \, \sin (3 \, \beta) - \sin (\beta) \right) \,
\sin (2 \, \gamma) \,\mathrm{Re} (\kappa_{6}) \right]
\end{eqnarray}
\begin{eqnarray}
\frac{\partial V_{min}}{\partial \gamma} &=& 2 \, v^2 \, \sin^2 (\beta)\,
\left[ 2 \, \sqrt{3} \, u \, v_0 \, \cos(\alpha) \, \sin (2 \, \gamma) \,
  y_1 \right.
\\ \nonumber &-& \left. \sqrt{3} \, u^2 \, \sin (2 \, \alpha) \, \sin
(2 \, \gamma) \, y_2 \right. \\ \nonumber &+& \left. v^2 \, \sin ^2
(\beta) \, \sin (4 \,
\gamma) \, y_3 \right. \\ \nonumber &+& \left. 6
\, v \, v_0 \, \cos (\beta) \, \cos (2 \, \gamma) \, \mathrm{Re}(\kappa_{6}) \right]
\end{eqnarray}
\end{subequations}
with $y_i$ being defined as:
\begin{subequations}
\begin{gather}
y_1 = \mathrm{Re} (\omega_{2} + \omega_{3} - \omega_{1}) \\
y_2 = 2 \, \mathrm{Re}(\kappa_{1} - \kappa_{2}) +
\tilde{\kappa}_{1}- \tilde{\kappa}_{2} + \tau_{3} \\
y_3 = \lambda_{3} ^{\xi} - 3 \, \lambda_{2} ^{\xi}
\end{gather}
\end{subequations}

\noindent All three equations \Eqref{mincond} have to be equal zero. A
more restrictive requirement is that all the terms should vanish
separately. The $y_i$, $\mathrm{Re} (\kappa_{6})$ and
$\mathrm{Re}(\sigma_{3})$ are parameters of the Higgs potential
and should not be constrained to vanish in order to avoid
accidental symmetries arising in the potential. Therefore their
coefficients should vanish separately. This poses restrictions on
the angles $\alpha$, $\beta$ and $\gamma$ as well as on the moduli $v$, $u$
and $v_0$. Obviously, one of the solutions is given by $u=0$
($\alpha$ is then no longer a variable), $\beta= \arccos
(1/\sqrt{3})$ and $\gamma= \pi/4$ , i.e. the VEVs of the fields
$\phi_{1,2}$ vanish, the VEVs of the fields $\xi_{i}$ are equal
$\frac{v}{\sqrt{3}}$ and $\phi_{0}$ has an in general
non-vanishing VEV $v_0$. Assuming that all quartic couplings are
of the same order and all mass parameters have the same order, it
is natural that the VEV for the $\xi_{i}$ fields is of the order
of $v_0$. This means only a slight parameter tuning is necessary
to achieve the equivalence of these VEVs as is needed for the
zeroth order approximation of the fermion masses in our first
numerical example.  It is noteworthy that the number of (massless)
Goldstone bosons is increased by two, if one additionally sets the
parameters $y_{i}$, $\mathrm{Re} (\sigma_{3})$ and $\mathrm{Re}
(\kappa _{6})$ to zero, see
\Appref{app:Higgsspectrum1}. I.e. the restrictive
requirement that none of them vanishes turns out to be sufficient to
avoid further Goldstone bosons. In the numerical study it is pointed
out that an auxiliary \Groupname{Z}{2} symmetry can explain the
required Yukawa couplings. This \Groupname{Z}{2}, if also valid in the
Higgs sector (and therefore in the whole Lagrangian), restricts the
quartic couplings. It enforces $\sigma_{3}$, $\kappa_{3}$,
$\kappa_{4}$, $\omega_{1}$, $\omega_{2}$ and $\omega_{3}$ to
vanish. As far as we can see this does not create an accidental
symmetry in the Higgs potential and so does not change the
discussion.

\noindent A similar analysis can be done for our second numerical
example which enforces the VEVs of $\phi_{0}$, $\phi_{2}$ and
$\xi_{1}$ to be equal and the other VEVs to vanish in the zeroth
approximation. This corresponds to $\alpha= \frac{\pi}{2}$ and
$\beta=0$ ($\gamma$ is then irrelevant). One sees that also in
this case the coefficients of the parameters $y_i$, $\mathrm{Re} (\kappa_{6})$ and
$\mathrm{Re}(\sigma_{3})$ vanish such that the VEV conditions
\Eqref{mincond} can be fulfilled. Note that the values of $v_{0}$, $u$
and $v$ are not constrained to be equal, but again it is plausible
that they are nearly the same, if the mass parameters $\mu_{i}$ as
well as the quartic couplings of the Higgs potential are chosen to be
of similar size. Interestingly, setting the parameters
$y_{i}$,  $\mathrm{Re} (\kappa_{6})$ and
$\mathrm{Re}(\sigma_{3})$ to zero increases the symmetry of the
potential also at this minimum and leads to three further Goldstone
bosons (see \Appref{app:Higgsspectrum2}) which is actually one more
than in the case above. By inspecting the Higgs potential $V$ one
finds that at least for real VEVs and parameters
of the potential requiring $y_{i}=0$, $\mathrm{Re} (\sigma_{3})=0$ and
$\mathrm{Re} (\kappa_{6})=0$ causes an accidental $SO(2) _{acc}$ symmetry under
which $\left( \phi_{1}, \phi _{2} \right) ^{T}$ forms a doublet and the
other fields remain invariant and an accidental $SO(3) _{acc}$ under which
$\left( \xi_{1}, \xi_{2}, \xi_{3} \right) ^{T}$ transforms as triplet
and the fields $\phi_{0}$, $\phi_{1}$, $\phi_{2}$ trivially. The VEV
configuration with vanishing $\langle \phi_{1,2} \rangle$ then only
breaks $SO(3) _{acc}$ and not $SO(2) _{acc}$ and therefore gives rise to two
Goldstone bosons whereas the configuration $\langle \phi_{0} \rangle
\neq 0$, $\langle \phi_{2} \rangle \neq 0$ and $\langle \xi_{1}
\rangle \neq 0$ breaks both accidental symmetries resulting in three
Goldstone bosons.

\noindent In the limiting case that all
mass parameters $\mu _{i}$ of the potential are equal, these two
minima are exactly degenerate (along with many others), since the
value of the potential at the minima is $-\frac{1}{2} \, (\mu_{1} ^{2}
\, v_{0} ^{2} + \mu_{3} ^{2} \, v^{2})$ and $-\frac{1}{2} \, (\mu_{1}
^{2} \, v_{0} ^{2} + \mu_{2} ^{2} \, u^{2} + \mu_{3} ^{2} \, v^{2})$, respectively.

\noindent Further investigation of the minima with CP-violation
which lead to realistic masses and mixing parameters is beyond the
scope of the paper, but it is plausible that these minima can
be formed through small deformations starting with CP-conserving minima, as done here.

\noindent We did not perform any checks of the stability of each
minimum and the potential as a whole, since the number of parameters
($\sim 30$) makes us confident that there exists at least one point in
the parameter space for each minimum where it fulfills together with
the potential all the stability criteria. Furthermore we did not
address the question whether the minimum is a
local or global one, since this might also only depend on an
appropriate choice of the parameters.

\section{Conclusion and Outlook}

To conclude, we have presented a low energy model based on the SM
gauge group augmented with the flavor symmetry \Groupname{S}{4}.
In contrast to other flavor models we used the requirement to
embed our model into a GUT like $SO(10)$ and at the same time into
a continuous flavor group like $SO(3) _{f}$ or $SU(3) _{f}$ as
guideline for the choice of the transformation properties of
fermion and Higgs fields under \Groupname{S}{4}. The resulting
model is minimal in that sense. 

\noindent Since the structure of the mass matrices is
determined by \Groupname{S}{4}, it is not obvious whether we can
accommodate all observed masses and mixing angles, even though the
model contains as many parameters as observables needed to be
fixed in the CP-conserving case.

\noindent To check this, we explore two cases which are
perturbations around two different rank one mass textures for
quarks and charged leptons that can be maintained for two choices of ground
states of the theory together with some tuning of the Yukawa
couplings. The first is the so-called democratic mass
matrix and the second one only has a non-vanishing 2-3 block. We
give numerical examples for each that are able to fit the known
fermion masses and mixing angles in the quark and lepton sector up
to rather small deviations. We believe that invoking radiative
corrections will lead to full accordance with the experimental
data. The difference between the mixings of quarks and leptons
crucially depends on the fact that the form of the mass matrix of
the light neutrinos differs strongly from the one of the quarks
and charged leptons. In our first example an auxiliary
\Groupname{Z}{2} can help to explain this difference and in the second one
the parameters have to be fine-tuned. Taking the auxiliary
\Groupname{Z}{2} as an exact symmetry of the theory prevents the
model from being embedded into $SO(10)$, since the right-handed
neutrinos transform differently from the other fermions. However,
one can still promote our model to an $SU(5)$ GUT. As the Higgs
fields $\phi_{1,2}$ transform under \Groupname{Z}{2} whereas
$\xi_{1,2,3}$ remain invariant, the \Groupname{Z}{2} is not
compatible with any embedding of our \Groupname{S}{4} flavor
symmetry into $SO(3) _{f}$ and $SU(3) _{f}$ without adding further
fields.

\noindent The right-handed neutrinos are degenerate at tree-level
and even more their mass matrix is proportional to the unit
matrix. Therefore the large leptonic mixing angles have to be
encoded in the structure of the Dirac mass matrices for the
neutrinos and charged leptons. 

\noindent The VEV configurations we used in our
numerical examples can only be analyzed in the CP-conserving limit,
since the Higgs potential turns out to be quite complicated.
Nevertheless these are determined by our flavor symmetry. In contrast
to this, the values of the Yukawa couplings are not fixed by
\Groupname{S}{4}. The question
why the top quark is 36 times heavier than the bottom quark remains
unanswered, but can possibly be explained, if our model is promoted to
$SO(10) \times G_{f}$.

\noindent Throughout this paper we have not been concerned with the
question how to guarantee that the Higgs spectrum just contains one
light uncharged (scalar) Higgs inducing only flavor diagonal
interactions like the Higgs in the SM while the rest is
heavier. Connected to this is the problem of flavor changing neutral
currents and lepton flavor
violations which usually arise in models with more than one SM-like
Higgs doublet. Typically these effects are negligible, if the masses
of the flavor changing Higgs fields are above a few $\TeV$.
Systematic calculations are difficult, since
the Higgs mass spectrum cannot be evaluated in general cases.

\noindent Finally, we want to comment on the possibility of
supersymmetrizing  our model. Introducing supersymmetry apart from
its salient feature to solve the hierarchy problem if it is broken
at low energies technically leads to a severe simplification of
the Higgs potential, since then all the quartic terms are
determined by the D-terms. The danger lies in the fact that this
generally leads to large accidental global symmetries in the
potential which consequently lead to a number of unwanted
Goldstone bosons. Two ways of treating this problem
can  be found in the literature: a.)
breaking the discrete and hence also the accidental symmetries by
the soft SUSY breaking terms (for example: \cite{susy1}) or b.)
introduce gauge singlets whose couplings are invariant under the
discrete symmetry, but break the accidental ones (for example:
\cite{susy2}). Obviously, the whole situation can change in a grand
unified model, since then the Higgs doublet fields can belong to
various representations of the GUT which have different invariant
couplings (see for example \cite{susyso10couplings} for a SUSY
$SO(10)$ model). Since supersymmetric potentials are restricted to be
positive by construction, checks of their stability are easier than
for non-supersymmetric ones. These issues are currently under study.

\subsubsection*{Acknowledgements}
The work of R.N.M.~was supported by the National Science Foundation
grant no.\ Phy--0354401 and the
Alexander von Humboldt Foundation (the Humboldt Research Award).
This work was supported by
the ``Sonderforschungsbereich~375 f\"ur Astro-{}Teil\-chen\-phy\-sik der
Deutschen Forschungsgemeinschaft''.

\appendix

\small
\mathversion{bold}
\section{Appendix: Group Theory of \Groupname{S}{4}}
\mathversion{normal}
\label{app:grouptheory}
In this appendix we display the representation matrices, Kronecker
products and Clebsch Gordan coefficients to calculate all the terms being invariant under the group \Groupname{S}{4}.

\subsection{Representation Matrices}
\label{app:repmats}
The representation matrices fulfilling \Eqref{generatoreq} can be chosen as:
\[\rm A=\left(\begin{array}{cc}
                                                -1 & 0 \\
                                                 0 & 1
                                \end{array}\right)
                       \;\;\; \mbox{and} \;\;\;
\rm B=-\frac{1}{2} \left(\begin{array}{cc}
                                                1   & \sqrt{3} \\
                                                -\sqrt{3} & 1
                                \end{array}\right)
\;\;\; \mbox{for} \;\;\; \Rep{2} \; ,
\]
\[\rm A=\left(\begin{array}{ccc}
                                                -1 & 0 & 0 \\
                                                 0 & 0 & - 1 \\
                                                 0 & 1 & 0
                                \end{array}\right)
                        \;\;\; \mbox{and} \;\;\;
\rm B=\left(\begin{array}{ccc}
                                                0 & 0 & 1 \\
                                                 1 & 0 & 0 \\
                                                 0 & 1 & 0
                                \end{array}\right)
\;\;\; \mbox{for} \;\;\; \MoreRep{3}{1}
\]
and
\[\rm A=\left(\begin{array}{ccc}
                                                1 & 0 & 0 \\
                                                 0 & 0 & 1 \\
                                                 0 & - 1 & 0
                                \end{array}\right)
                       \;\;\; \mbox{and} \;\;\;
\rm B=\left(\begin{array}{ccc}
                                                0 & 0 & 1 \\
                                                 1 & 0 & 0 \\
                                                 0 & 1 & 0
                                \end{array}\right)
\;\;\; \mbox{for} \;\;\; \MoreRep{3}{2} \; .
\]
These matrices can be found in \cite{Lomont}.

\subsection{Kronecker Products}
\label{app:kronprods}
The Kronecker products can be calculated from the above given
character table \cite{Hamermesh}.

\parbox{3in}{
\begin{eqnarray}\nonumber
\MoreRep{1}{i} \times \MoreRep{1}{j} &=& \MoreRep{1}{\scriptsize \rm
  (i+j) mod 2 +1 \normalsize}  \;\;\; \forall \; \rm i \; \mbox{and}
\; \rm j \\ \nonumber
\Rep{2} \times \MoreRep{1}{i} &=& \Rep{2}  \;\;\; \forall \; \rm i \\ \nonumber
\MoreRep{3}{i} \times \MoreRep{1}{j} &=& \MoreRep{3}{\scriptsize \rm
  (i+j) mod 2 +1 \normalsize}  \;\;\; \forall \; \rm i \; \mbox{and}
\; \rm j
\end{eqnarray}}
\parbox{3in}{
\begin{eqnarray}\nonumber
\MoreRep{3}{i} \times \Rep{2} &=&  \MoreRep{3}{1} +
  \MoreRep{3}{2}  \;\;\; \forall \; \rm i \\ \nonumber
\MoreRep{3}{1} \times \MoreRep{3}{2} &=& \MoreRep{1}{2} + \Rep{2} +
\MoreRep{3}{1} + \MoreRep{3}{2}
\end{eqnarray}}
\[
\left[ \Rep{2} \times \Rep{2}\right] = \MoreRep{1}{1} + \Rep{2} \; ,
\;\;\;\; \left\{ \Rep{2} \times \Rep{2}\right\} = \MoreRep{1}{2}
\;\;\; \mbox{and} \;\;\;
\left[ \MoreRep{3}{i} \times \MoreRep{3}{i}\right] = \MoreRep{1}{1} + \Rep{2} + \MoreRep{3}{1} \; , \;\;\;\; \left\{ \MoreRep{3}{i} \times \MoreRep{3}{i}\right\} = \MoreRep{3}{2}  \;\;\; \forall \; \rm i
\]
where we introduced the notation $\left[\mu \times \mu \right]$
for the symmetric and $\left\{ \mu \times \mu
\right\}$ for the anti-symmetric part of the product $\mu \times
\mu$. 

\noindent Note that $\nu \times \mu = \mu \times \nu$ for all representations
$\mu$ and $\nu$.

\subsection{Clebsch Gordan Coefficients}
\label{app:CGcoeffs}
The Clebsch Gordan coefficients can be calculated \cite{Cornwell} with the given
representation matrices for
\[
A, A^{\prime} \sim \MoreRep{1}{1} \;\;\; , \;\;\; B,B^{\prime} \sim \MoreRep{1}{2} \;\;\;
, \;\;\; \left( \begin{array}{cc} a_{1} \\ a_{2} \end{array} \right),
 \left( \begin{array}{cc} a^{\prime}_{1} \\ a^{\prime}_{2} \end{array} \right)
  \sim \Rep{2} \; , \;\;
\left( \begin{array}{ccc} b_{1}\\ b_{2} \\ b_{3} \end{array}
\right), \left( \begin{array}{ccc} b_{1} ^{\prime}\\ b^{\prime}_{2} \\ b^{\prime}_{3} \end{array}
\right) \sim \MoreRep{3}{1} \;\;\; \mbox{and} \] \[
  \left( \begin{array}{ccc} c_{1}\\ c_{2} \\ c_{3} \end{array}
\right), \left( \begin{array}{ccc} c^{\prime}_{1}\\ c^{\prime}_{2} \\ c^{\prime}_{3} \end{array}
\right)  \sim \MoreRep{3}{2} \; .
\]
Since we choose all the representation matrices to be real, it also holds:
\[
A^{\star} \sim \MoreRep{1}{1} \;\;\; , \;\;\; B^{\star} \sim \MoreRep{1}{2} \;\;\;
, \;\;\; \left( \begin{array}{cc} a_{1}^{\star} \\ a_{2} ^{\star}
  \end{array} \right) \sim \Rep{2} \; , \;\;
\left( \begin{array}{ccc} b_{1} ^{\star}\\ b_{2}^{\star} \\ b_{3}^{\star} \end{array}
\right) \sim \MoreRep{3}{1} \;\;\; \mbox{and} \;\;\;  \left(
  \begin{array}{ccc} c_{1} ^{\star} \\ c_{2} ^{\star} \\ c_{3} ^{\star} \end{array}
\right) \sim \MoreRep{3}{2} \; .
\]

\noindent The Clebsch Gordan coefficients for the one-dimensional
representations are trivial:
\[
A \, A^{\prime} \sim \MoreRep{1}{1} \;\;\; , \;\;\; A \, B \sim
\MoreRep{1}{2} \;\;\; , \;\;\; B \, A \sim \MoreRep{1}{2} \;\;\; , \;\;\; B \, B^{\prime} \sim \MoreRep{1}{1}
\]
as well as the products $\MoreRep{1}{1} \times \mu$ of any
representation $\mu$ with the total singlet
$\MoreRep{1}{1}$:
\[
\left( \begin{array}{c} A \, a_1 \\ A \, a_2 \end{array} \right) \sim
\Rep{2} \;\;\; , \;\;\; \left( \begin{array}{c} A \, b_1 \\ A
    \, b_2 \\ A \, b_3 \end{array} \right) \sim \MoreRep{3}{1}  \;\;\;
, \;\;\; \left( \begin{array}{c} A \, c_1 \\ A \, c_2 \\ A \, c_3 \end{array} \right) \sim \MoreRep{3}{2} \; .
\]
And here are the ones for $\MoreRep{1}{2} \times \mu$ of any
representation $\mu$:
\[
\left( \begin{array}{c} -B \, a_2 \\ B \, a_1 \end{array} \right) \sim
\Rep{2} \;\;\; , \;\;\; \left( \begin{array}{c} B \, b_1 \\ B
    \, b_2 \\ B \, b_3 \end{array} \right) \sim \MoreRep{3}{2}  \;\;\;
, \;\;\; \left( \begin{array}{c} B \, c_1 \\ B \, c_2 \\ B \, c_3 \end{array} \right) \sim \MoreRep{3}{1} \; .
\]
The Clebsch Gordan coefficients for $\mu \times \mu$ have the form:\\
\parbox{2.5in}{\begin{center}
 for  $\Rep{2}$
\begin{eqnarray}\nonumber
&a_1 a^{\prime}_1 + a_2 a^{\prime}_2 \sim \MoreRep{1}{1}&\\ \nonumber
&-a_1 a^{\prime}_2 + a_2 a^{\prime}_1 \sim \MoreRep{1}{2}&\\ \nonumber
&\left( \begin{array}{c} a_1 a^{\prime}_2 + a_2 a^{\prime}_1 \\ a_1 a^{\prime}_1 - a_2 a^{\prime}_2 \end{array} \right) \sim \Rep{2}&
\end{eqnarray}
\end{center}
}
\parbox{4in}{\begin{center}
for $\MoreRep{3}{1}$
\begin{eqnarray}\nonumber
&\sum \limits _{j=1} ^{3} b_j b^{\prime}_j  \sim \MoreRep{1}{1}&\\ \nonumber
&\left( \begin{array}{c} \frac{1}{\sqrt{2}} (b_2 b^{\prime}
_2 - b_3 b^{\prime}_3) \\ \frac{1}{\sqrt{6}} (-2 b_1 b^{\prime}_1 + b_2 b^{\prime}_2 + b_3 b^{\prime}_3) \end{array} \right) \sim \Rep{2}& \\ \nonumber
&\left( \begin{array}{c} b_2 b^{\prime}_3 + b_3 b^{\prime}_2 \\ b_1 b^{\prime}_3 + b_3 b^{\prime}_1\\ b_1
    b^{\prime}_2 + b_2 b^{\prime}_1 \end{array} \right) \sim
\MoreRep{3}{1} \; , \;\; \left(
  \begin{array}{c} b_3 b^{\prime}_2 - b_2 b^{\prime}_3 \\ b_1 b^{\prime}_3 - b_3 b^{\prime}_1 \\ b_2 b^{\prime}_1 -
  b_1 b^{\prime}_2 \end{array} \right) \sim \MoreRep{3}{2}&
\end{eqnarray}
\end{center}}
\parbox{3.5in}{\begin{center}
for $\MoreRep{3}{2}$
\begin{eqnarray} \nonumber
&\sum \limits _{j=1} ^{3} c_j c^{\prime}_j \sim \MoreRep{1}{1}&\\ \nonumber
&\left( \begin{array}{c} \frac{1}{\sqrt{2}} (c_2 c^{\prime}
_2 - c_3 c^{\prime}_3) \\ \frac{1}{\sqrt{6}} (-2 c_1 c^{\prime}_1 + c_2 c^{\prime}_2 + c_3 c^{\prime}_3) \end{array} \right) \sim \Rep{2}& \\ \nonumber
&\left( \begin{array}{c} c_2 c^{\prime}_3 + c_3 c^{\prime}_2 \\ c_1 c^{\prime}_3 + c_3 c^{\prime}_1\\ c_1
    c^{\prime}_2 + c_2 c^{\prime}_1 \end{array} \right) \sim \MoreRep{3}{1}\; , \;\;\left(
  \begin{array}{c} c_3 c^{\prime}_2 - c_2 c^{\prime}_3 \\ c_1 c^{\prime}_3 - c_3 c^{\prime}_1 \\ c_2 c^{\prime}_1 -
  c_1 c^{\prime}_2 \end{array} \right) \sim \MoreRep{3}{2}& .
\end{eqnarray}
\end{center}}\\
Note here that the parts belonging to the symmetric part of the
product $\mu \times \mu$ are symmetric under the interchange of
unprimed and primed whereas the ones belonging to the anti-symmetric
part change sign, i.e. are anti-symmetric. 

\noindent Note also that for our
choice of generators the Clebsch
Gordan coefficients for $\MoreRep{3}{1} \times \MoreRep{3}{1}$ and
$\MoreRep{3}{2} \times \MoreRep{3}{2}$ turn out to be the same.
For the coupling $\Rep{2} \times \MoreRep{3}{1}$ the Clebsch Gordan
coefficients are
\parbox{2.5in}{
\begin{eqnarray} \nonumber
\left( \begin{array}{c} a_2 b_1 \\ -\frac{1}{2}(\sqrt{3} a_1 b_2 + a_2
    b_2)\\  \frac{1}{2} (\sqrt{3} a_1 b_3 - a_2 b_3) \end{array}
\right) \sim \MoreRep{3}{1}\\ \nonumber
\left( \begin{array}{c} a_1 b_1 \\ \frac{1}{2}(\sqrt{3} a_2 b_2 - a_1
    b_2)\\  -\frac{1}{2} (\sqrt{3} a_2 b_3 + a_1 b_3) \end{array}
\right) \sim \MoreRep{3}{2}
\end{eqnarray}}
\parbox{1in}{and for  $\Rep{2} \times \MoreRep{3}{2}$}
\parbox{2in}{
\begin{eqnarray} \nonumber
\left( \begin{array}{c} a_1 c_1 \\ \frac{1}{2}(\sqrt{3} a_2 c_2 - a_1
    c_2)\\  -\frac{1}{2} (\sqrt{3} a_2 c_3 + a_1 c_3) \end{array}
\right) \sim \MoreRep{3}{1}\\ \nonumber
\left( \begin{array}{c} a_2 c_1 \\ -\frac{1}{2}(\sqrt{3} a_1 c_2 + a_2
    c_2)\\  \frac{1}{2} (\sqrt{3} a_1 c_3 - a_2 c_3) \end{array}
\right) \sim \MoreRep{3}{2} .
\end{eqnarray}}
\begin{center}
\hspace{-2.8in} And for $\MoreRep{3}{1} \times \MoreRep{3}{2}$ one finds the following
combinations:
\begin{eqnarray} \nonumber
& \sum \limits _{j=1} ^{3} b_j c_j \sim \MoreRep{1}{2}&\\ \nonumber
&\left( \begin{array}{c}  \frac{1}{\sqrt{6}} (2 b_1 c_1 - b_2 c_2 - b_3
    c_3) \\ \frac{1}{\sqrt{2}} (b_2 c
_2 - b_3 c_3) \end{array} \right) \sim \Rep{2}& \\ \nonumber
&\left( \begin{array}{c} b_3 c_2 - b_2 c_3 \\ b_1 c_3 - b_3 c_1 \\ b_2 c_1 -
  b_1 c_2 \end{array} \right) \sim \MoreRep{3}{1}\; , \;\;\left(
  \begin{array}{c} b_2 c_3 + b_3 c_2 \\ b_1 c_3 + b_3 c_1\\ b_1
    c_2 + b_2 c_1 \end{array} \right) \sim \MoreRep{3}{2}& .
\end{eqnarray}
\end{center}

\mathversion{bold}
\subsection{Embeddings of \Groupname{S}{4} into $SO(3)$ and $SU(3)$}
\mathversion{normal}
\label{app:embeds}
We only display the resolution of the smallest representations of
$SO(3)$ ($SU(3)$) into irreducible ones of \Groupname{S}{4}.

\parbox{2.5in}{
\begin{eqnarray} \nonumber
&\underline{SO(3)}& \;\;\; \rightarrow \;\;\; \underline{\mbox{\Groupname{S}{4}}}\\ \nonumber
&\Rep{1}& \;\;\; \rightarrow \;\;\; \MoreRep{1}{1} \\ \nonumber
&\Rep{3}& \;\;\; \rightarrow \;\;\; \MoreRep{3}{2} \\ \nonumber
&\Rep{5}& \;\;\; \rightarrow \;\;\; \Rep{2} + \MoreRep{3}{1} \\ \nonumber
&\Rep{7}& \;\;\; \rightarrow \;\;\; \MoreRep{1}{2} + \MoreRep{3}{1} +
\MoreRep{3}{2} \\ \nonumber
&\Rep{9}& \;\;\; \rightarrow \;\;\; \MoreRep{1}{1} + \Rep{2} +
\MoreRep{3}{1} + \MoreRep{3}{2}
\end{eqnarray}}
\parbox{2.5in}{
\begin{eqnarray} \nonumber
&\underline{SU(3)}& \;\;\; \rightarrow \;\;\; \underline{\mbox{\Groupname{S}{4}}}\\ \nonumber
&\Rep{1}& \;\;\; \rightarrow \;\;\; \MoreRep{1}{1} \\ \nonumber
&\Rep{3}& \;\;\; \rightarrow \;\;\; \MoreRep{3}{2} \\ \nonumber
&\Rep{6}& \;\;\; \rightarrow \;\;\; \MoreRep{1}{1} + \Rep{2} + \MoreRep{3}{1}\\ \nonumber
&\Rep{8}& \;\;\; \rightarrow \;\;\; \Rep{2} + \MoreRep{3}{1} + \MoreRep{3}{2} \\ \nonumber
&\Rep{10}& \;\;\; \rightarrow \;\;\; \MoreRep{1}{2} + \MoreRep{3}{1} + 2 \; \MoreRep{3}{2}
\end{eqnarray}}\\
\noindent The first table can be found in \cite{Lomont} and the second one can
be calculated with the formulae given in \cite{Patera}.

\section{Appendix: Minimization of the Higgs Potential}
\label{app:higgspotential}
\subsection{Remaining VEV Conditions}
\label{app:remainmincond}
The derivatives $\frac{\partial V_{min}}{\partial v}$,
$\frac{\partial V_{min}}{\partial u}$ and  $\frac{\partial
  V_{min}}{\partial v_0}$ have the following form:
\small
\begin{eqnarray} \nonumber
\frac{\partial V_{min}}{\partial v} &=& 2 \, v \, \left(- \mu_3 ^2 + 2
  \, (\lambda_1 ^{\xi} + 4 \, \lambda_2 ^{\xi}) \, v^2 \right) + 2 \,
v \, v_0^2 \, (2 \, \mathrm{Re}(\kappa_5) + \tilde{\kappa}_{5} +
\tau_1) \\ \nonumber &+& 2 \, v \, u^2 \, (4 \, \mathrm{Re}(\kappa_1 + \kappa_2) + 2
\, (\tilde{\kappa}_1 + \tilde{\kappa}_2) + \tau_2) + 9 \, v_0 \, v^2 \,
\mathrm{Re}(\kappa_6) \, \sin (\beta) \, \sin (2 \, \beta) \, \sin (2
\, \gamma) \\ \nonumber &+& 4 \, u \, v_0 \, v \, \left[ \sin(\alpha) \left( 2 -
      3 \, \sin^2(\beta)\right) - \sqrt{3} \, \cos (\alpha) \, \sin^2
    (\beta) \, \cos (2 \, \gamma) \right] \, y_1 \\ \nonumber &+& 2 \, u^2 \, v \,
  \left[ -\cos (2 \, \alpha) \, (2 - 3 \, \sin^2 (\beta)) + \sqrt{3} \,
    \sin (2 \, \alpha) \, \sin^2 (\beta) \, \cos (2 \, \gamma)
  \right] \, y_2 \\ \nonumber &+& 4 \, v^3 \, \left[ 4 \, \sin^2 (\beta) - (3 + \cos
    ^2 (2 \, \gamma)) \, \sin^4 (\beta) \right] \, y_3 \\ \nonumber
\frac{\partial V_{min}}{\partial v_0} &=& 2 \, v_0 \, \left( - \mu_1
  ^2 + 2 \, \lambda_0 \, v_0 ^2 \right) + 2 \, v_0 \, u^2 \, (\sigma_1 + 2 \,
\mathrm{Re}(\sigma_2) + \tilde{\sigma}_2) + 2 \, u^3 \,
\mathrm{Re}(\sigma_3) \, \sin (3 \, \alpha) \\ \nonumber &+& 2 \, v_0 \, v^2 \, (2 \,
\mathrm{Re}(\kappa_5) + \tilde{\kappa}_5 + \tau_1) + 3 \, v^3 \,
\mathrm{Re}(\kappa_6) \, \sin(\beta) \, \sin (2 \, \beta) \, \sin (2
\, \gamma) \\ \nonumber &+& 2 \, u \, v^2 \, \left[ \sin (\alpha) \, (2 - 3 \, \sin
  ^2 (\beta)) - \sqrt{3} \, \cos (\alpha) \, \sin^2 (\beta) \, \cos (2
  \, \gamma) \right] \, y_1\\ \nonumber
\frac{\partial V_{min}}{\partial u} &=& 2 \, u \, \left( - \mu_2 ^2 + 2 \,
(\lambda_1 + \lambda_3) \, u^2 \right) + 2 \, u \, v_0 ^2 \, (\sigma_1 +2 \,
\mathrm{Re}(\sigma_2) + \tilde{\sigma}_2) + 6 \, v_0 \, u^2 \,
\mathrm{Re}(\sigma_3) \, \sin(3 \, \alpha) \\ \nonumber &+& 2 \, (4 \,
\mathrm{Re}(\kappa_1 + \kappa_2) + 2 \, (\tilde{\kappa}_1 +
\tilde{\kappa}_2) + \tau_2) \, u \, v^2 \\ \nonumber &+& 2 \, v_0 \, v^2 \, \left[ \sin (\alpha)
  \, (2 - 3 \, \sin^2 (\beta)) - \sqrt{3} \, \cos(\alpha) \, \sin^2
  (\beta) \, \cos(2 \, \gamma) \right] \, y_1 \\ \nonumber &+&  2 \, u \, v^2 \,
\left[ -\cos(2 \, \alpha) \, (2 - 3 \, \sin^2 (\beta)) + \sqrt{3} \,
  \sin(2 \, \alpha) \, \sin^2 (\beta) \, \cos (2 \, \gamma) \right] \,
y_2
\end{eqnarray}
\normalsize

\noindent In the following sections we present the Higgs mass matrices
$\mathcal{M} ^{2}$ for the two minima around which we have perturbed
to find our numerical solutions shown above. 
\noindent We use the following parameterization for the SM-like Higgs doublets
$\phi$:
\[
\phi = \left( \begin{array}{c}
    \mbox{\footnotesize{VEV}} + \phi ^{r} + i \; \phi ^{i} \\
    \phi ^{c \, r} + i \; \phi ^{c \, i}
\end{array} \right) \; .
\]
We define $\mathcal{M} ^{2}$ as:
\[
\mathcal{M} ^{2} = \left. \frac{\partial ^{2} V}{\partial \phi ^{\mathrm{x}}
  \, \partial \tilde{\phi} ^{\mathrm{x}}} \right| _{\footnotesize \mbox{all fields = 0}}
\]
\normalsize
where $\phi, \tilde{\phi} \in \left\{ \phi_{0}, \phi_{1}, \phi_{2}, \xi_{1},
  \xi_{2}, \xi_{3} \right\}$ and $\mathrm{x} \in \left\{ r, i , c  \,
  r , c \, i \right\}$. \\
\noindent We give the mass matrices in the basis $\left\{ \xi_{1}
  ^{\mathrm{x}}, \xi_{2} ^{\mathrm{x}},
    \xi_{3} ^{\mathrm{x}} , \phi_{1} ^{\mathrm{x}} , \phi_{2}
    ^{\mathrm{x}} , \phi_{0} ^{\mathrm{x}} \right\}$ where
  $\mathrm{x}= r,i, c\, r, c\, i$ for the different components of the
  Higgs doublet fields.
\noindent For the calculation of the mass matrices we have assumed that all the
parameters in the Higgs potential are real such that there is no
mixing between the real and the imaginary parts of the components of the Higgs doublet fields.

\mathversion{bold}
\subsection{Mass Spectrum for the Minimum $\langle \xi_{i} \rangle =
  \frac{v}{\sqrt{3}}$, $\langle \phi_{1,2}\rangle =0$ and $\langle \phi_{0} \rangle=v_0$}
\mathversion{normal}
\label{app:Higgsspectrum1}

\noindent The non-trivial VEV conditions in this case are:
\small
\begin{eqnarray} \nonumber
v \, (2 \, \sqrt{3} \, v \, v_0 \, \mathrm{Re}(\kappa_{6}) + \frac{2}{3} \, (3 \,
\lambda_{1} ^{\xi} + 4 \, \lambda_{3} ^{\xi}) \, v^2 - \mu_{3} ^2 + (2
\, \mathrm{Re}(\kappa_{5}) + \tilde{\kappa}_{5} + \tau_{1}) \, v_0 ^2)&=&0\\ \nonumber
\frac{2}{\sqrt{3}} \, \mathrm{Re}(\kappa_{6}) \, v^3 + 2 \, \lambda_0
\, v_0 ^3 - v_0 \, \mu_1 ^2 + v^2 \, v_0 \, (2 \,\mathrm{Re}(\kappa_5)
+ \tilde{\kappa}_5 + \tau_1)&=&0
\end{eqnarray}
\normalsize

\noindent The resulting mass matrices for the Higgs fields have the following
structure:
\small
\[
\mathcal{M} ^{2}= \left( \begin{array}{cccccc}
   m_1 & m_2 & m_2 & 0 & -2 \, m_6 & m_3\\
   . & m_1 & m_2 & \sqrt{3} \, m_6 & m_6 & m_3 \\
   . & . & m_1 & - \sqrt{3} \, m_6 & m_6 & m_3 \\
   . & . & . & m_4 & 0 & 0\\
   . & . & . & .& m_4 & 0\\
   . & . & . & . & . & m_5
\end{array} \right)
\]
\normalsize
\noindent The eigenvalues of such a matrix are given by:
\small
\begin{eqnarray}\nonumber
\frac{1}{2} \, \left( m_1 + 2 \, m_2 + m_5 \pm \sqrt{(m_1 + 2 \, m_2 -
  m_5)^2 + 12 \, m_3 ^2} \right) \\ \nonumber
\frac{1}{2} \, \left( m_1 - m_2 + m_4 \pm \sqrt{(m_1 - m_2 - m_4)^2 +
    24 \, m_6 ^2} \right) \;\;\; \mbox{{each two times}}
\end{eqnarray}
\normalsize
\noindent The corresponding characteristic polynomial is given by:
\small
\[
\left[ -6 \, m_6 ^2 + \left(m_1 - m_2 - \chi \right) \left( m_4 - \chi
    \right) \right]^2 \left[
  -3 \, m_3 ^2 + \left( m_1 + 2 \, m_2 -\chi \right) \left( m_5 - \chi
    \right) \right]=0
\]
\normalsize
\noindent For the fields $\phi ^{c \, r}$ and $\phi ^{c \, i}$ the variables
$m_{i}$ with $i=1,...,6$ have the following form:
\small
\begin{eqnarray}\nonumber
m_1 &=&  2 \, (2 \, v^2 \, \lambda_{1} ^{\xi} - \mu_{3} ^{2} + v_0 ^2 \, \tau_{1}) \\ \nonumber
m_2 &=&  \frac{4}{3} \, v \, (\sqrt{3} \, v_0 \, \mathrm{Re}
(\kappa_{6}) + 2 \, \lambda_{3} ^{\xi} \, v)\\ \nonumber
m_3 &=& \frac{2}{3} \, v \, (\sqrt{3} \, v_0 \, (2 \, \mathrm{Re} (\kappa_{5}) + \tilde{\kappa}
_{5})+ 2 \, \mathrm{Re}(\kappa_{6}) \, v ) \\ \nonumber
m_4 &=& 2 \, (- \mu _{2} ^{2} + v_0^2 \, \sigma_{1} + \tau_{2} \, v^2 )\\ \nonumber
m_5 &=& 2 \, ( 2 \, \lambda_{0} \, v_0^2 - \mu_{1} ^2 + \tau_{1} \,
v^2) \\ \nonumber
m_6 &=& - \frac{2}{3} \, v \, (2 \, v \, \mathrm{Re}(\kappa_{3})+ \sqrt{3} \, v_0 \,
\mathrm{Re} (\omega_{2} + \omega_{3}))
\end{eqnarray}
\normalsize
\noindent For the uncharged scalar fields $\phi ^{r}$:
\small
\begin{eqnarray} \nonumber
m_1 &=& 2 \, (\frac{2}{3} \, v^2 \, (5 \, \lambda _{1} ^{\xi} + 8 \,
\lambda_{2} ^{\xi} + 4 \, \lambda_{3} ^{\xi}) - \mu_{3} ^{2} + v _{0}
^{2} \, (2 \, \mathrm{Re} (\kappa_{5}) + \tilde{\kappa} _{5} + \tau_{1})) \\ \nonumber
m_2 &=& \frac{4}{3} \, v \, (3 \, \sqrt{3} \,
\mathrm{Re}(\kappa_{6}) \, v_0 + 2 \, v \, (\lambda_{1} ^{\xi} - 2 \,
\lambda_{2} ^{\xi} + 2 \, \lambda_{3} ^{\xi}) )\\ \nonumber
m_3 &=& \frac{4}{3} \, v \, (3 \, \mathrm{Re}(\kappa_{6}) \, v + \sqrt{3} \, (2 \,
\mathrm{Re}(\kappa_{5}) + \tilde{\kappa}_{5} + \tau_{1}) \, v_0
)\\ \nonumber
m_4 &=& 2 \, ( -\mu_{2} ^2 + (\sigma_{1} + 2 \,
\mathrm{Re}(\sigma_{2}) + \tilde{\sigma}_{2}) \, v_0 ^{2} + ( 4 \,
\mathrm{Re}(\kappa_{1} + \kappa_{2}) + 2 \, (\tilde{\kappa}_{1} +
\tilde{\kappa}_{2}) + \tau_{2}) \, v^2)\\ \nonumber
m_5 &=& 2 \, (6 \, \lambda_{0} \, v_0^2 - \mu_1 ^2 + (2 \,
\mathrm{Re}(\kappa_{5}) + \tilde{\kappa}_{5} + \tau_1) \, v^2) \\ \nonumber
m_6 &=& \frac{4}{\sqrt{3}} \, v_0 \, v \, \mathrm{Re}(\omega_1 -\omega_2 - \omega_3)
\end{eqnarray}
\normalsize
\noindent and for the uncharged pseudo-scalars  $\phi ^{i}$:
\small
\begin{eqnarray} \nonumber
m_1 &=& \frac{4}{3} \, v^2 \, (3 \, \lambda_{1} ^{\xi}  - 4 \, \lambda_{4} ^{\xi}) - 2 \, \mu_{3} ^{2} + 2 \, v_{0} ^{2} \, (\tilde{\kappa}_{5} + \tau_{1} - 2 \, \mathrm{Re} (\kappa_{5})) \\ \nonumber
m_2 &=& \frac{4}{3} \, v \, (\sqrt{3} \, \mathrm{Re}(\kappa_6) \, v_0 + 2
\, (\lambda_{3} ^{\xi} + \lambda_{4} ^{\xi}) \, v)\\ \nonumber
m_3 &=& \frac{4}{3} \, v \, (2 \, \sqrt{3} \, v_0 \, \mathrm{Re}(\kappa_{5}) +
\mathrm{Re}(\kappa_{6}) \, v) \\ \nonumber
m_4 &=& 2 \,( - \mu_2 ^2 + (\sigma_{1} - 2 \, \mathrm{Re}(\sigma_{2})
+ \tilde{\sigma}_{2}) \,
v_0 ^2 - (4 \, \mathrm{Re}(\kappa_1 + \kappa_2) - 2 \,
(\tilde{\kappa}_1 + \tilde{\kappa}_2) - \tau_{2}) \, v^2)\\ \nonumber
m_5 &=& 2 \, ( 2 \, \lambda_{0} \, v_0 ^2 - \mu_1 ^2 +
(\tilde{\kappa}_{5} + \tau_1 -2 \, \mathrm{Re}(\kappa_5)) \, v^2) \\ \nonumber
m_6 &=& - \frac{4}{3} \,
v \, (\mathrm{Re}(\kappa_3 - \sqrt{3} \, \kappa_4) \, v + \sqrt{3} \, \mathrm{Re}(\omega_2) \, v_0)
\end{eqnarray}
\normalsize

\mathversion{bold}
\subsection{Mass Spectrum for the Minimum $\langle \phi_{0} \rangle=v_0$, $\langle \phi_{2} \rangle = u$, $\langle \xi_{1}\rangle = v$ and $\langle \xi_{2,3}\rangle =\langle \phi_{1} \rangle=0$}
\mathversion{normal}
\label{app:Higgsspectrum2}
\noindent The three non-trivial VEV conditions are:
\small
\begin{eqnarray} \nonumber
v \, (2 \, v^2 \, (\lambda_{1} ^{\xi} + 4 \, \lambda_{2} ^{\xi})
- \mu_{3} ^{2} + v_{0} ^2 \, (2 \, \mathrm{Re} (\kappa_{5}) +
\tilde{\kappa} _{5} + \tau_{1}) + u^2 \, (4 \, (2 \, \mathrm{Re}
(\kappa_{1}) + \tilde{\kappa} _{1}) + \tau _{2} + 2 \, \tau_{3}) &
 & \\ \nonumber + 4 \, u \, v_{0} \, \mathrm{Re} (\omega_{2} + \omega_{3} - \omega_{1})) &=& 0 \\ \nonumber
2 \, u^3 \, (\lambda_{1} + \lambda_{3})- 3 \, u^2 \, v_{0} \,
\mathrm{Re}(\sigma_{3}) -\mu_{2} ^{2} \, u + v_0^2 \, u \, (\sigma_{1}
+ 2 \, \mathrm{Re} (\sigma_{2}) + \tilde{\sigma}_{2})  & & \\
\nonumber + u \, v^2 \, (4 \, (2 \, \mathrm{Re} (\kappa_{1}) + \tilde{\kappa} _{1}) + \tau _{2} + 2
\, \tau_{3})  + 2 \, v^2 \, v_{0} \, \mathrm{Re} (\omega_{2} + \omega_{3} - \omega_{1}) &=& 0 \\ \nonumber
2 \, v_{0} ^3 \, \lambda _{0} - \mu_{1} ^{2} \, v_{0} + v_{0}
\, u^2 \, (\sigma_{1} + 2 \, \mathrm{Re} (\sigma_{2}) + \tilde{\sigma}
_{2}) - u^3 \, \mathrm{Re} (\sigma_{3}) + v^2  \, v_{0} \, (2 \,
\mathrm{Re} (\kappa_{5}) + \tilde{\kappa} _{5} + \tau_{1}) & & \\ \nonumber + 2  \, v^{2} \, u \, \mathrm{Re} (\omega_{2} + \omega_{3} - \omega_{1}) &=& 0
\end{eqnarray}
\normalsize
\noindent The mass matrices for the Higgs scalars have a block structure:
\small
\[
\mathcal{M} ^{2}= \left( \begin{array}{cccccc}
        m_{1} &  0 & 0 & 0 & m_{8} & m_{9} \\
        .     &  m_{2} & m_{3} & 0 & 0 & 0 \\
        .  & . & m_{2} & 0 & 0 & 0 \\
        . & . & . & m_{4} & 0 & 0 \\
        . & . & . & .& m_{5} & m_{7}\\
        . & . & . & . & . & m_{6}
\end{array} \right)
\]
\normalsize
The corresponding eigenvalues are:
\small
\[
m_{4} \; , \;\; m_{2} \pm m_{3}
\]
\normalsize
\noindent and the solutions of the characteristic polynomial:
\small
\[
\left|   \begin{array}{ccc}
                        m_{1} - \chi & m_{8} & m_{9}\\
                        m_{8} & m_{5} - \chi & m_{7}\\
                        m_{9} & m_{7} & m_{6} - \chi
                 \end{array}
\right| = 0
\]
\normalsize
\noindent For the fields $\phi ^{c \, r}$ and $\phi ^{c \, i}$ the variables
$m_{i}$ with $i=1,...,9$ have the following form:
\small
\begin{eqnarray}\nonumber
m_1 &=&  2 \, (2 \, v^2 \, (\lambda _{1} ^{\xi} + 4 \, \lambda_{2}
^{\xi}) - \mu_{3} ^{2} + v_{0} ^{2} \, \tau_{1} + u^2 \, (\tau_{2} + 2
\, \tau_{3}) - 4 \, u \, v_{0} \, \mathrm{Re} (\omega _{1})) \\ \nonumber
m_2 &=&  2 \, (2 \, v^2 \, (\lambda_{1} ^{\xi} - 2 \, \lambda _{2}
^{\xi}) - \mu_{3} ^{2} + v_{0} ^{2} \, \tau_{1} + u^{2} \, (\tau_{2} -
\tau_{3}) + 2 \, u \, v_{0} \, \mathrm{Re} (\omega_{1}))\\ \nonumber
m_3 &=&  4 \, v \, (2 \, u \ \mathrm{Re} (\kappa_{3}) + v_{0} \,
\mathrm{Re} (\kappa_{6}))\\ \nonumber
m_4 &=&  2 \, (2 \, u^2 \, (\lambda_{1} - \lambda_{3}) - \mu_{2} ^{2}
+ v_{0} ^{2} \, \sigma_{1} + v^{2} \, (\tau_{2} - 2 \, \tau_{3}) + 2
\, u \, v_{0} \, \mathrm{Re} (\sigma_{3}))\\ \nonumber
m_5 &=&  2 \, (2 \, u^2 \, (\lambda_{1} + \lambda_{3}) - \mu_{2} ^{2}
+ v_{0} ^{2} \, \sigma_{1} + v^{2} \, (\tau_{2} + 2 \, \tau_{3}) -2 \,
u \, v_{0} \, \mathrm{Re} (\sigma_{3}))\\ \nonumber
m_6 &=&  2 \, (2 \, v_{0} ^{2} \, \lambda_{0} - \mu_{1} ^{2} + u^{2}
\, \sigma_{1} + v^{2} \, \tau_{1})\\ \nonumber
m_7 &=&  2 \, (u \, v_{0} \, (2 \, \mathrm{Re} (\sigma_{2}) +
\tilde{\sigma} _{2}) - u^{2} \, \mathrm{Re} (\sigma_{3}) - 2 \, v^{2}
\, \mathrm{Re} (\omega_{1}))\\ \nonumber
m_8 &=& 4 \, v \, (2 \, u \, (2 \, \mathrm{Re} (\kappa_{1}) +
\tilde{\kappa} _{1}) + v_{0} \, \mathrm{Re} (\omega_{2} + \omega_{3})) \\ \nonumber
m_9 &=& 2 \, v \, (v_{0} \, (2 \, \mathrm{Re} (\kappa_{5}) +
\tilde{\kappa} _{5}) + 2 \, u \, \mathrm{Re} (\omega_{2} + \omega_{3}))
\end{eqnarray}
\normalsize
\noindent For the uncharged scalar fields $\phi ^{r}$:
\small
\begin{eqnarray} \nonumber
m_1 &=& 2 \, (6 \, v^{2} \, (\lambda_{1} ^{\xi} + 4 \, \lambda_{2}
^{\xi}) - \mu_{3} ^{2} + v_{0}^{2} \, (2 \, \mathrm{Re} (\kappa _{5})
+ \tilde{\kappa} _{5} + \tau_{1}) + u^{2} \, (4 \, ( 2 \,\mathrm{Re}
(\kappa_{1}) + \tilde{\kappa} _{1}) + \tau_{2} + 2 \, \tau_{3})  \\
\nonumber & & + 4 \, u \, v_{0} \, \mathrm{Re} (\omega_{2} + \omega_{3}-\omega_{1}))\\ \nonumber
m_2 &=& 2 \, (2 \, v^{2} \, (\lambda_{1} ^{\xi} - 2 \, \lambda_{2}
^{\xi} + 2 \, \lambda _{3} ^{\xi}) - \mu_{3} ^{2} + v_{0} ^{2} \, (2
\, \mathrm{Re} (\kappa_{5}) + \tilde{\kappa} _{5} + \tau_{1}) + u^{2}
\, (2 \, \mathrm{Re} (\kappa _{1}) \\ \nonumber & & + \tilde{\kappa} _{1} + 3 \, (2 \,
\mathrm{Re} (\kappa _{2}) + \tilde{\kappa} _{2}) + \tau_{2} -
\tau_{3}) + 2 \, u \, v_{0} \, \mathrm{Re} (\omega_{1} - \omega_{2} - \omega_{3}))\\ \nonumber
m_3 &=& 12 \, v \, v_{0} \, \mathrm{Re} (\kappa_{6})\\ \nonumber
m_4 &=& 2 \, (2 \, u^{2} \, (\lambda_{1} + \lambda_{3}) - \mu_{2} ^{2}
+ v_{0} ^{2} \, (\sigma_{1} + 2 \, \mathrm{Re} (\sigma_{2}) +
\tilde{\sigma} _{2}) + v^{2} \, (4 \, ( 2 \, \mathrm{Re} (\kappa_{2})
+ \tilde{\kappa} _{2}) + \tau_{2} - 2 \, \tau_{3}) \\ \nonumber & & + 6 \, u \, v_{0} \,
\mathrm{Re} (\sigma_{3}))\\ \nonumber
m_5 &=& 2 \, (6 \, u^{2} \ (\lambda_{1} + \lambda_{3}) - \mu_{2} ^{2}
+ v_{0} ^{2} \, (\sigma_{1} + 2 \, \mathrm{Re} (\sigma_{2}) +
\tilde{\sigma} _{2}) + v^{2} \, (4 \, ( 2 \,  \mathrm{Re} (\kappa
_{1}) + \tilde{\kappa} _{1}) + \tau_{2} + 2 \, \tau_{3}) \\ \nonumber & & - 6 \, u \, v_{0} \,
\mathrm{Re} (\sigma_{3}))\\ \nonumber
m_6 &=& 2 \, (6 \, v_{0} ^{2} \, \lambda _{0} - \mu_{1} ^{2} + u^{2}
\, (\sigma_{1} + 2 \, \mathrm{Re} (\sigma_{2}) + \tilde{\sigma}_{2}) +
v^{2} \, (2 \, \mathrm{Re} (\kappa_{5}) + \tilde{\kappa} _{5} + \tau_{1}))\\ \nonumber
m_7 &=& 2 \, (2 \, v^{2} \, \mathrm{Re} (\omega_{2} + \omega_{3} -
\omega_{1}) + 2 \, u \, v_{0} \, (\sigma_{1} + 2 \, \mathrm{Re}
(\sigma_{2}) + \tilde{\sigma} _{2}) - 3 \, u^{2} \, \mathrm{Re} (\sigma_{3}))\\ \nonumber
m_8 &=& 4 \, v \, (u \, (4 \, ( 2 \, \mathrm{Re} (\kappa _{1}) +
\tilde{\kappa} _{1}) + \tau_{2} + 2 \, \tau_{3}) + 2 \, v_{0} \,
\mathrm{Re} (\omega_{2} + \omega_{3} - \omega_{1}))\\ \nonumber
m_9 &=& 4 \, v \, (v_{0} \, (2 \, \mathrm{Re} (\kappa _{5}) +
\tilde{\kappa} _{5} + \tau_{1}) + 2 \, u \, \mathrm{Re} (\omega_{2} +
\omega_{3} - \omega_{1}))
\end{eqnarray}
\normalsize
\noindent and for the uncharged pseudo-scalars  $\phi ^{i}$:
\small
\begin{eqnarray} \nonumber
m_1 &=& 2 \, (2 \, v^{2} \, (\lambda _{1} ^{\xi} + 4 \, \lambda _{2}
^{\xi}) - \mu_{3} ^{2} + v_{0} ^{2} \, (\tilde{\kappa} _{5}
- 2 \, \mathrm{Re} (\kappa _{5})  + \tau_{1}) + u^{2} \, (4 \, (\tilde{\kappa} _{1}
- 2 \, \mathrm{Re} (\kappa _{1})) + \tau_{2} + 2 \, \tau_{3}) \\
\nonumber & & - 4 \, u \, v_{0} \, \mathrm{Re} (\omega_{1} + \omega_{2} - \omega_{3}))\\ \nonumber
m_2 &=& -2 \, (2 \, v^{2} \, (- \lambda_{1} ^{\xi} + 2 \, \lambda _{2}
^{\xi} + 2 \, \lambda _{4} ^{\xi}) + \mu_{3} ^{2} + v_{0} ^{2} \, (2
\, \mathrm{Re} (\kappa_{5}) - \tilde{\kappa} _{5} - \tau_{1}) + u^{2}
\, (2 \, \mathrm{Re} (\kappa_{1}) \\ \nonumber & & - \tilde{\kappa}_{1} + 3 \, (2 \,
\mathrm{Re} (\kappa_{2}) - \tilde{\kappa} _{2}) -\tau_{2} + \tau_{3})
- 2 \, u \, v_{0} \, \mathrm{Re} (\omega_{1} + \omega_{2} - \omega_{3}))\\ \nonumber
m_3 &=& 4 \, v \, (2 \, u \, \mathrm{Re}(\kappa _{3} - \sqrt{3} \,
\kappa _{4}) + v_{0} \, \mathrm{Re} (\kappa _{6}))\\ \nonumber
m_4 &=& 2 \, (2 \, u^{2} \, (\lambda_{1} - 2 \, \lambda_{2} -
\lambda_{3}) - \mu_{2} ^{2} + v_{0} ^{2} \, (\sigma_{1} - 2 \,
\mathrm{Re} (\sigma_{2}) + \tilde{\sigma} _{2}) + v^{2} \, (4 \,
(\tilde{\kappa} _{2} - 2 \, \mathrm{Re} (\kappa_{2})) + \tau_{2} - 2 \,
\tau_{3}) \\ \nonumber & & + 2 \, u \, v_{0} \, \mathrm{Re} (\sigma _{3}))\\ \nonumber
m_5 &=& 2 \, (2 \, u^{2} \, (\lambda_{1} + \lambda_{3}) - \mu_{2} ^{2}
+ v_{0} ^{2} \, (\sigma_{1} - 2 \, \mathrm{Re} (\sigma_{2}) +
\tilde{\sigma} _{2}) + v^{2} \, (4 \, (\tilde{\kappa}  _{1} - 2 \,
\mathrm{Re} (\kappa_{1})) + \tau_{2} + 2 \, \tau_{3}) \\ \nonumber & & - 2 \, u \,
v_{0} \, \mathrm{Re} (\sigma_{3}))\\ \nonumber
m_6 &=& 2 \, (2 \, v_{0} ^{2} \, \lambda _{0} - \mu_{1} ^{2} + u^{2}
\, (\sigma_{1} - 2 \, \mathrm{Re}(\sigma_{2}) + \tilde{\sigma} _{2}) +
v^{2} \, (\tilde{\kappa} _{5} - 2 \, \mathrm{Re} (\kappa _{5})  + \tau_{1}))\\ \nonumber
m_7 &=& -2 \, (u^{2} \, \mathrm{Re} (\sigma_{3}) + 2 \, v^{2} \,
\mathrm{Re} (\omega_{1} + \omega_{2} - \omega_{3}) - 4 \, u \, v_{0}
\, \mathrm{Re} (\sigma_{2}))\\ \nonumber
m_8 &=& 8 \, v \,  (4 \, u \, \mathrm{Re} (\kappa _{1}) + v _{0} \,
\mathrm{Re} (\omega _{2}))\\ \nonumber
m_9 &=& 8 \, v \, (v_{0} \, \mathrm{Re} (\kappa _{5}) + u \,
\mathrm{Re} (\omega _{2}))
\end{eqnarray}

\end{document}